\documentclass[prd,twocolumn,superscriptaddress,altaffilletter,amssymb,showpacs,nofootinbib,notitlepage,10pt]{revtex4}

\usepackage{graphicx}
\usepackage{amsmath}
\usepackage{epsfig}
\usepackage{bm}

\newcommand{\be}{\begin{equation}}
\newcommand{\ee}{\end{equation}}
\newcommand{\beq}{\begin{eqnarray}}
\newcommand{\eeq}{\end{eqnarray}}

\begin{document}

\title{Thawing models in the presence of a generalized Chaplygin gas}

\author{Sergio del Campo }
 \email{sdelcamp@ucv.cl}
\affiliation{\it Instituto de F\'{\i}sica, Pontificia Universidad
de Cat\'{o}lica de Valpara\'{\i}so, Casilla 4950, Valpara\'{\i}so,
Chile}

\author{Carlos R. Fadragas}
\email{fadragas@uclv.edu.cu}
\affiliation{\it Departamento de F\'isica, Universidad Central de Las
Villas, 54830 Santa Clara, Cuba.}

\author{Ram\'{o}n Herrera }
 \email{ramon.herrera@ucv.cl}
\affiliation{\it Instituto de F\'{\i}sica, Pontificia Universidad
de Cat\'{o}lica de Valpara\'{\i}so, Casilla 4950, Valpara\'{\i}so,
Chile}

\author{Carlos Leiva }
 \email{cleivas@uta.cl}

\affiliation{\it Departamento de F\'{\i}sica, Universidad de
Tarapac\'{a}, Casilla 7-D, Arica, Chile}

 \author{Genly Leon}
\email{genly.leon@ucv.cl}
\affiliation{\it Instituto de F\'{\i}sica, Pontificia Universidad
de Cat\'{o}lica de Valpara\'{\i}so, Casilla 4950, Valpara\'{\i}so,
Chile}

\author{Joel Saavedra}
 \email{Joel.Saavedra@ucv.cl }
 \affiliation{\it Instituto de F\'{\i}sica, Pontificia Universidad
de Cat\'{o}lica de Valpara\'{\i}so, Casilla 4950, Valpara\'{\i}so,
Chile}

\date{\today}

\begin{abstract}

In this paper we consider a cosmological model whose main components are a scalar field and a generalized Chaplygin gas. We obtain an exact solution for a flat arbitrary potential. This solution have the right dust limit when  the Chaplygin parameter  $A\rightarrow 0$. We use the dynamical systems approach in order to describe  the cosmological evolution of the mixture for an exponential self-interacting scalar field potential. We study the scalar field with an arbitrary self-interacting potential using the ``Method of $f$-devisers.'' Our results are illustrated for the special case of a coshlike potential.  We find  that usual scalar-field-dominated and scaling solutions  cannot be late-time attractors in the presence of the Chaplygin gas (with $\alpha>0$). We recover the standard results at the dust limit ($A\rightarrow 0$). In particular, for the exponential potential, the late-time attractor is a pure generalized Chaplygin solution mimicking an effective cosmological constant.  In the case of arbitrary potentials, the late-time attractors are de Sitter solutions in the form of a cosmological constant, a pure generalized Chaplygin solution or a continuum of solutions, when the scalar field and the Chaplygin gas densities are of the same orders of magnitude. The different situations depend on the parameter choices.
\end{abstract}



\maketitle

\section*{Introduction}\label{sec:Int}

It is  common knowledge that the expansion of the Universe is currently passing through an accelerated phase \cite{R98,P99}. All the observational data from these former references until the current measurements of redshift and
luminosity-distance relations of type Ia Supernovae
(SNe)\cite{S10} are in agreement with this accelerated era. These observations are indicating the presence of a vacuum energy, the well-known cosmological constant \cite{Peebles:2002gy,S03,T04}. Another possible description  is the existence of the scalar field that is evolving in a universe described by a Friedmann-Robertson-Walker (FRW) geometry, the quintessence model \cite{Zlatev:1998tr}. The current acceleration of the universe is one of the important problems of  modern cosmology. This problem appears in  Einstein's standard general relativity, and one of the proposals to solve it, within this framework, is to consider one exotic component in the matter content of the universe
\cite{Gabadadze:2007dv}, the so-called dark energy component (for a review about this, see Ref. \cite{Copeland:2006wr}). We would like to focus our attention on models where dark energy is described by a homogeneous scalar field ($\phi(t)$) \cite{scalarD-E}. In order to describe the dynamics of  dark energy (quintessence), the equation of state parameter (EOS), $\omega_\phi\equiv
p_\phi/\rho_\phi$ plays a crucial role. Its current value is close to $-1$. Therefore, the cosmic evolution of dark energy through the scalar field is described by the barotropic parameter $\omega_\phi$. In  Ref. \cite{Caldwell:2005tm}, the authors distinguished between two categories, the freezing ($d\omega_\phi/d\phi<0$) and thawing ($d\omega_\phi/d\phi>0$) models. These quintessence models are characterized by a scalar field potential that asymptotically goes to zero \cite{Scherrer:2007pu}. There are several references about this kind of cosmological model (for a summary see  Refs. \cite{thawing-M}). On the other hand, Chaplygin gas models have been widely investigated in the literature \cite{Bento:2002ps,Bilic:2001cg,Bento:2002yx,Gorini:2002kf,Debnath:2004cd,Zhu:2004aq,Zhang:2004gc,Chakraborty:2007ui,Sen:2005sk,Barreiro:2004bd,Ali:2011sv,Fabris:2010yh,delCampo:2009cz}. The modified Chaplygin gas for the $k=0$ FRW universe is exactly the same as adding a bulk viscosity proportional to a power of the fluid density \cite{Barrow:1988yc, Barrow:1990vx}. The generalized Chaplygin gas (GCG) has been investigated
from the dynamical systems viewpoint, for example, in \cite{Bhadra:2011ac,Mazumder:2011ke,Li:2008uv,He:2008zzc,Rudra:2011ku}.

In this paper we would like to extend the analysis in \cite{Scherrer:2007pu} by considering a more general matter component, that is a  GCG and a scalar field characterized by its self-interacting scalar potential. We obtain an exact solution for a flat arbitrary potential, that have the right dust limit when the Chaplygin parameter $A \rightarrow 0$ \cite{Scherrer:2007pu}. In order to motivate the analysis for a general
(arbitrary) potential, we first consider the simple case of an exponential potential $V(\phi)=V_0 e^{-\lambda \phi}$ \cite{Copeland:1997et,Halliwell:1986ja,Barreiro:1999zs,Burd:1988ss,Sami:2002fs,Rubano:2001su,Heard:2002dr,Coley:1997nk,Liddle:1988tb,Rubano:2003et,Guo:2003eu, Barrow:1994nt, Aguirregabiria:1993pm,Aguirregabiria:1993pk,Ibanez:1995zs,Pavluchenko:2003ge,Goheer:2002ac,Piedipalumbo:2011bj,Fang:2006zq,Aguirregabiria:1996uh,Copeland:2009be}.  Then, we study an arbitrary self-interacting scalar field potential. In this case we use the ``Method of $f$-devisers'' presented in \cite{Escobar:2013js}. This method allows to perform the phase space analysis without specifying the potentials \emph{ad initium}, and then one just substitute the desired forms, instead of repeating the whole procedure for every distinct potential. The method is a refinement of a method that has been applied to isotropic (FRW)
scenarios \cite{Copeland:2009be,Fang:2008fw,Matos:2009hf,Leyva:2009zz, UrenaLopez:2011ur,Dutta:2009yb}, and that has been generalized to several cosmological contexts \cite{Escobar:2012cq,Escobar:2011cz, Farajollahi:2011ym, Xiao:2011nh}.

This article is organized as follow, in section \ref{sec:Thawing} we present the cosmological model under consideration. Section \ref{sec:II} is devoted to the study of the exponential potential an the corresponding dynamical system. In section \ref{newmethod} we study the phase space of the cosmological model for general potential $V(\phi)$. Finally, we conclude in the section \ref{Final_rem}.

\section{The cosmological model}\label{sec:Thawing}
In this section we would like to describe a cosmological setting
compose by a minimally-coupled scalar field  that describes the dark
energy and a Chaplygin gas that behaves as dark matter in the
appropriate limit.

The cosmological equations are given by
 \begin{eqnarray}
 H^2-\frac{\rho_{ch}+\rho_\phi}{3}&=&0, \label{ecH}\\
\dot{\rho_{ch}}+3H(\rho_{ch}+P_{ch})&=&0,  \label{ecch} \\
\dot{\rho}_\phi+3H(\rho_\phi+P_\phi)&=&0, \label{ecfi} \\
\nonumber
\end{eqnarray}
where we work in units in which $8 \pi G=1$, $H$ is the Hubble
constant, $\rho_{ch}$ and $\rho_{\phi}$ are the Chaplygin gas and scalar field densities, respectively. For simplicity, the radiation component is neglected. $P_{ch}$ and
$P_{\phi}$ represent the Chaplygin gas and scalar field pressures,
respectively. For the scalar field we have:
\begin{eqnarray}
\rho_\phi=\frac{\dot{\phi}^2}{2}+V(\phi),  \\
P_\phi=\frac{\dot{\phi}^2}{2}-V(\phi).
\end{eqnarray}
On the other hand, the EoS for the GCG is
\begin{equation}
P_{ch}=-\frac{A}{\rho_{ch}^\alpha}, \label{ecstatech}
\end{equation}
{ where $A$ is a positive constant and $\alpha$ is a constant
with an upper bound, $\alpha\leq 1$. 
In particular,  when $\alpha=1$ corresponds to the original
Chaplygin gas.}
 In the framework of FRW cosmology, this EoS
leads, after inserted into the relativistic energy conservation
equation, to an evolution of the energy density as
\begin{equation}
\rho_{ch}=\left(A+\frac{B}{a^{3(\alpha+1)}}\right)^{\frac{1}{\alpha
+1}}
=\rho_{ch0}\left[B_s+\frac{(1-B_s)}{a^{3(\alpha+1)}}\right]^{\frac{1}{\alpha+1}}.\label{roch}
\end{equation}
{Here, $a$ is the scale factor and B is a positive integration
constant.  In this way, the GCG is characterized by two
parameters, $B_s=A/\rho_{ch0}^{1+\alpha}$ and $\alpha$. Here,
$\rho_{ch 0}$ is the current value of $\rho_{ch}$, considering
that $a=1$ at the present. These parameter has been confronted with the 
observational data, see Refs.\cite{const,const1}. In particular,
the values of $B_s=0.73_{-0.06}^{+0.06}$ and
$\alpha=-0.09_{-0.12}^{+0.15}$ were obtained in Ref.\cite{const1}.
Also, in Ref.\cite{Bento:2002ps} the values $0.81\lesssim
B_s\lesssim 0.85$ and $0.2\lesssim\alpha\lesssim 0.6$ were found
from the observational data arising from different colaborations, such that Archeops 
(by using the first peak localization)  and BOOMERANG (by using the third peak localization).
Recently, the values 
$B_s=0.775_{-0.0161-0.0338}^{+0.0161+0.037}$ and
$\alpha=0.00126_{-0.00126-0.00126}^{+0.000970+0.00268}$ were
obtained from Markov Chain Monte Carlo method \cite{const2}.}
For the phase space simulations implemented in the present paper we select the value $\alpha=0.5,$ first in agreement with Archeops and BOOMERANG collaborations, and second, following the reference \cite{DelPopolo:2013bpa}. This value seems to be large in comparison with the  observational values in \cite{const2}, however, if we consider large values for $\alpha$, and include the effect of shear and rotation, then when studying the evolution of the perturbations in GCG universes, it is found that that the joint effect of shear and rotation is that of slowing down the collapse with respect to the simple spherical collapse model. The described effect allows to solve the instability problems of the so-called unified dark matter models at the linear perturbation level \cite{DelPopolo:2013bpa}.

The evolution of the energy density $\rho_{ch}$ shows 
the behaviors of GCG at different times. At early times, the
energy density behaves as matter while at late times it behaves
like a cosmological constant. Then, this GCG in principle
describes both dark matter and dark energy in a single matter
component.

Now we can
define the barotropic index $\omega_{ch}$:
\begin{equation}\label{chapomega}
\omega_{ch}=-\frac{A}{A+\frac{B}{a^{3(\alpha+1)}}}.
\end{equation}
At late times, the universe is dominated by the scalar field and the GCG, neglecting the radiation
component.

Equations (\ref{ecch})
and (\ref{ecfi}) can be  rewritten in terms of the auxiliary
variables $x$, $y$ and $s$, defined by
\begin{eqnarray}
x&=&\frac{\phi'}{\sqrt{6}}, \nonumber \\
y&=&\sqrt{\frac{V(\phi)}{3H^2}},\label{aux1}\\
s&=&-\frac{1}{V}\,\frac{dV}{d\phi},\nonumber
\end{eqnarray}
where the prime denotes derivatives with respect to $\tau=\ln a$. Considering that the contribution for the kinetic and potential
energy are given by $x^2$ and $y^2$ respectively, the density
parameter of scalar field is given by
\begin{equation}\label{Omegaphi}
\Omega_\phi=\frac{\rho_\phi}{3H^2}=x^2+y^2,
\end{equation}
therefore the equation of state is
\begin{equation}\label{lsf}
 \gamma=1+\omega=\frac{2x^2}{x^2+y^2}.
\end{equation}

Thus, inserting the auxiliary (\ref{aux1}) into the equations of
motions (\ref{ecH}),(\ref{ecch}) and (\ref{ecfi}) we arrive  to the
following system
\begin{align}
&x'=-3x-\sqrt{\frac{3}{2}}s y^2+\frac{3x}{2}\{2x^2+(1-x^2-y^2)(1+\omega_{ch})\}, \label{eqx} \\
&y'=-\sqrt{\frac{3}{2}}s\,xy+\frac{3y}{2}\{2x^2+(1-x^2-y^2)(1+\omega_{ch})\},\label{eqy}\\
&s'=-\sqrt{6}\,s^2\,(\Gamma-1)x,\label{eqs}
\end{align}
where
\begin{equation}
\Gamma\equiv
V\,\left[\frac{dV}{d\,\phi}\right]^{-1} \frac{d^2 V}{d\,\phi^2}.
\end{equation}
Besides, the Friedmann constraint equation can be written as $\Omega_{\phi}+\Omega_{ch}=1$, and this implies $0 \leq \Omega_{\phi} \leq1$, for a non-negative density. Therefore the dynamical evolution of (\ref{eqx})-(\ref{eqs}) leave the coordinates $(x, y)$ within the upper-half unit disc.

Furthermore, the system \eqref{eqx}-\eqref{eqs} is non-autonomous since  \begin{equation}\label{omegach1}\omega_{ch}=-\frac{A}{A+{B}{e^{-3(\alpha+1)\tau}}},\end{equation} and it is in general not closed since $\Gamma$ does not depends a priori on the state variables $x,y,s.$

Before to perform a detailed analysis of the stability of this dynamical
system we rewrite it in terms of the observable quantities
$\Omega_{\phi}$ and $\gamma$, and the new equations are given by
\begin{eqnarray}
\gamma'&=&-3\gamma (2-\gamma )+s( 2-\gamma\ )\sqrt{ 3\gamma
\Omega_\phi },  \label{gama}\\
\Omega_\phi'&=&3(1-\gamma )\Omega_\phi (1-\Omega_\phi )+3\Omega_\phi
(1-\Omega_\phi )\omega_{ch}, \label{omegaprima}\\
s' &=& -\sqrt{3}\,s^2\,(\Gamma-1)\sqrt{\gamma \,
\Omega_\phi}.\label{lambda}
\end{eqnarray}

Equation \eqref{gama}  together with eq. (\ref{lambda}) encode the exact
description of the dynamic evolution of the scalar field. In any case to
find an exact solution it is a difficult task, and in order to proceed we consider two
assumptions. First, we consider that the barotropic parameter of
the scalar fluid is near to $-1$, therefore $\gamma\ll 1$. Second, we
consider a near flat potential that is $s$ is approximately
constant, say $s\approx s_0$  \cite{Scherrer:2007pu}.

In the limit  $\gamma\ll 1$ we obtain  from \eqref{omegaprima} an approximated equation with solution
\begin{equation}\label{Omega(a)}
\Omega_\phi (a)=\left[{\beta  \left(a^{-3 (\alpha +1)}+\chi \right)^{\frac{1}{\alpha +1}}+1}\right]^{-1}
\end{equation} satisfying $\Omega_\phi (a=1)=\Omega_{\phi 0},$ where $$\beta = {(1-\Omega_{\phi 0}) (\chi +1)^{-\frac{1}{\alpha +1}}}{\Omega_{\phi 0}}^{-1},$$ and
$\chi=A/B.$ Thus, in the limit $\chi\rightarrow 0$ is recovered the solution  described by the expression (25) in Ref. \cite{Scherrer:2007pu}: $\Omega_\phi=\left[1+\left(\Omega_{\phi 0}^{-1} -1\right)a^{-3}\right]^{-1},$ corresponding to standard quintessence.
From the solution \eqref{Omega(a)} and from equation \eqref{chapomega} we obtain the key formula
\begin{equation}\label{wchOmegaphi}
\omega_{ch}\left(\Omega_\phi\right)=-\chi  \beta ^{\alpha +1} (1-\Omega_\phi )^{-\alpha -1} \Omega_\phi ^{\alpha +1}.
\end{equation}

\begin{figure*}
\begin{center}
\begin{tabular}{cc}
\includegraphics[scale=0.8]{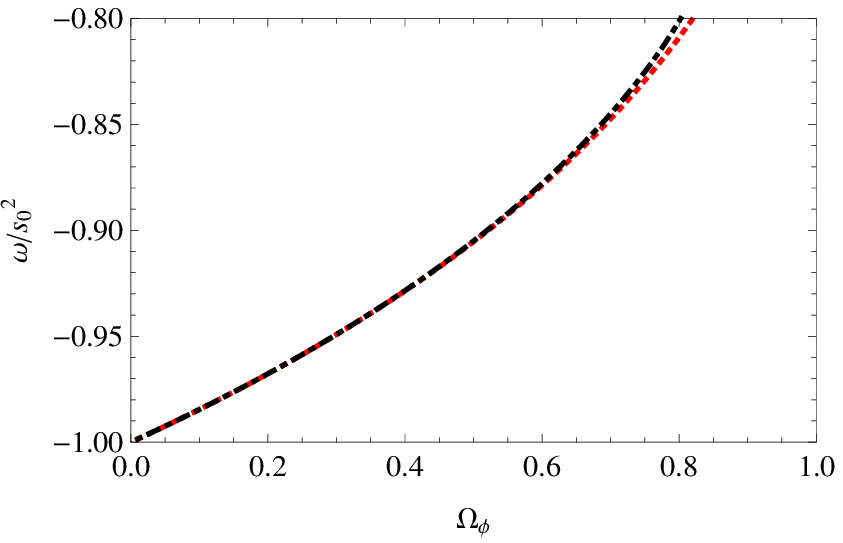} & \hspace{12pt} \includegraphics[scale=0.8]{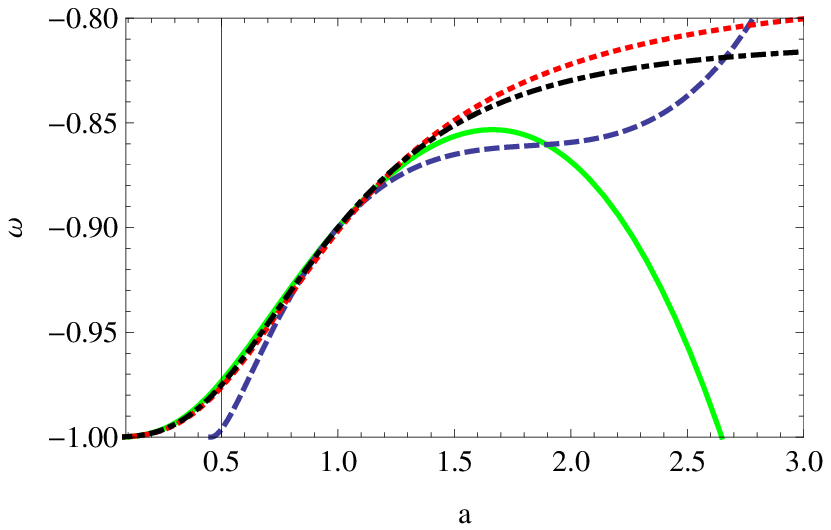}  \\
(a)& (b)
\end{tabular}
\caption{\label{fig1} {\it{
(a) The value of $\omega/s_{0}^2$ vs. $\Omega_\phi$ assuming a nearly flat potential and $\omega \sim -1.$ The dotted (red) line corresponds to the approximated solution discovered in \cite{Scherrer:2007pu} (equation (23) in \cite{Scherrer:2007pu}) and the dash-dotted (dark) line corresponds to the approximated solution, \eqref{eqfinal},  for $\chi=0.1,$  presented here.\\
(b) EoS parameter of the scalar field for the exponential potential with slope $s_0=0.8029.$
We have considered the initial conditions $\gamma(1)= 0.1, \Omega_\phi(1) = 0.7, s(1) =0.8029.$ The scale factor is normalized to 1 at present. The continuous (green) line corresponds to the exact value of $\omega(a)$ for model with solely a scalar field; the dashed (blue) line corresponds to the exact value of $\omega(a)$ for model that includes  the Chaplygin gas; the dotted (red) line corresponds to the approximated solution found in \cite{Scherrer:2007pu} (equation (23) in \cite{Scherrer:2007pu}) and the dash-dotted (dark) line corresponds to the approximated solution, \eqref{eqfinal}, presented here (assuming $\chi=0.1$).}}}
\end{center}
\end{figure*}
On the other hand, let us introduce the auxiliary function $\mu=\sqrt{\gamma}.$ Assuming that $\Omega_{\phi}$ is a monotonic function of the scale factor (in order to avoid that $d\Omega_\phi/d a=0$ at any value $a=a_0$), we obtain from \eqref{gama} and \eqref{omegaprima}
\begin{equation}
\frac{d \mu}{d\Omega_{\phi}}=-\frac{\left(\mu ^2-2\right) \left(3 \mu -\sqrt{3} s_0 \sqrt{\Omega_\phi }\right)}{6 (\Omega_\phi -1) \Omega_\phi  \left(-\mu ^2+\omega_{ch}+1\right)},
\end{equation}
where we have used the approximation $s\approx s_0=\text{const.}$. Using the hypothesis $\gamma\ll 1$ one is able to use the Taylor-expand   the above differential equation around $\mu=0$ up to second order to obtain the approximated equation
\begin{eqnarray}\label{approxOmegaprime}
\frac{d \mu}{d\Omega_{\phi}}&=&-\frac{s_0}{\sqrt{3} (\omega_{ch}+1) (\Omega_\phi -1) \sqrt{\Omega_\phi }}+\nonumber\\
&+&\frac{\mu }{(\omega_{ch}+1) (\Omega_\phi -1) \Omega_\phi }+{\cal O}\left(\mu \right)^2.
\end{eqnarray}
Substituting into the equation \eqref{approxOmegaprime} the expression $\omega_{ch}=\omega_{ch}(\Omega_{\phi})$ given by \eqref{wchOmegaphi} and integrating the resulting equation with the initial condition $\mu=0$ at $\Omega_{\phi}=0$ (which is true for the models we are considering here) we obtain the exact solution for $\alpha=1$ given by
 \begin{equation}
\mu(\Omega_\phi) =s_0\,{\cal F}(\Omega_\phi ) (F(\Omega |m)-E(\Omega |m))+\frac{s_0 \sqrt{\Omega_\phi }}{\sqrt{3}},
\end{equation} where
\begin{equation}{\cal F}(\Omega_\phi )=\frac{\sqrt{\Omega_\phi  \left(\beta  \sqrt{\chi }-1\right)+1} \sqrt{1-\Omega_\phi  \left(\beta  \sqrt{\chi }+1\right)}}{\Omega_\phi  \left(\beta  \sqrt{\chi }-1\right) \sqrt{3
   \beta  \sqrt{\chi }+3}},
\end{equation}
$F(\Omega|m)$ and $E(\Omega|m)$ are the elliptic integral of the first and second kind
respectively, with
\begin{equation}
\Omega=\sin ^{-1}\left(\sqrt{\Omega_\phi } \sqrt{\beta  \sqrt{\chi }+1}\right),
\end{equation} and
\begin{equation}
m=\frac{2}{\beta  \sqrt{\chi }+1}-1.
\end{equation}
Finally we obtain the expression
\begin{equation}\label{eqfinal}
1+\omega=\mu\left(\Omega_\phi\right)^2.
\end{equation}
Expression (\ref{eqfinal}) allows to obtain an exact
solution for the dynamical evolutions of the barotropic index and
using (\ref{Omega(a)}) we obtain $\omega=\omega(a)$. We would like
to note that our solutions have the right dust limit when
$A\rightarrow 0$ \cite{Scherrer:2007pu}, this behavior is shown in
Figures \ref{fig1} (a) and (b).

In fact, in the limit $A\rightarrow 0$ (i.e., $\chi\rightarrow
0$), the deviation between our solution \eqref{eqfinal} and the
solution (23) in \cite{Scherrer:2007pu} is given by the term
\begin{equation}
\frac{s_0^2 (\Omega_\phi -1) }{3 \Omega_\phi ^2}h(\Omega_\phi),
\end{equation} where
\begin{align}
& h(\Omega_\phi)=\left(\sqrt{\Omega_\phi }-E\left(\left.\sin ^{-1}\left(\sqrt{\Omega_\phi }\right)\right|1\right)\right) \times \nonumber \\ & \left(\sqrt{\Omega_\phi } (\Omega_\phi
   +1)+2 (\Omega_\phi -1) \tanh ^{-1}\left(\sqrt{\Omega_\phi }\right) +\right. \nonumber \\ & \left. -(\Omega_\phi -1) E\left(\left.\sin ^{-1}\left(\sqrt{\Omega_\phi }\right)\right|1\right)\right).
    \end{align}
    But
\begin{align}
& E\left(\Phi|m\right)=\int_0^{\Phi}\left[1-m\sin^2\theta\right]^{\frac{1}{2}}\mathrm{d}\theta\nonumber\\
& = \int_0^{\sin\Phi}\left[1-t^2\right]^{-\frac{1}{2}}\left[1-m t^2\right]^{\frac{1}{2}}\mathrm{d}t.
\end{align}
Thus,
\begin{align}
& E\left(\left.\sin ^{-1}\left(\sqrt{\Omega_\phi }\right)\right|1\right)=\int_0^{\sin ^{-1}\left(\sqrt{\Omega_\phi }\right)}\left[1-\sin^2\theta\right]^{\frac{1}{2}}\mathrm{d}\theta\nonumber\\
& = \int_0^{\sqrt{\Omega_\phi }}\mathrm{d}t=\sqrt{\Omega_\phi }.
\end{align}
This means that $h(\Omega_\phi)\equiv 0.$ That is, in the limit $A\rightarrow 0,$ our solution \eqref{eqfinal} and the solution (23) in \cite{Scherrer:2007pu} coincides.

Particularly, in the Figure \ref{fig1} (a) depict the value of $\omega/s_{0}^2$ vs. $\Omega_\phi$ assuming a nearly flat potential and $\omega \sim -1.$ The dotted (red) line corresponds to the approximated solution discovered in \cite{Scherrer:2007pu} (equation (23) in \cite{Scherrer:2007pu}) and the dash-dotted (dark) line corresponds to the approximated solution, \eqref{eqfinal},  for $\chi=0.1,$  presented here. In Figure \ref{fig1} (b) are displayed the EoS parameter of the scalar field for the exponential potential with constant slope. The scale factor is normalized to 1 at present. The continuous (green) line corresponds to the exact value of $\omega(a)$ for model with solely a scalar field; the dashed (blue) line corresponds to the exact value of $\omega(a)$ for model with the addition of the Chaplygin gas; the dotted (red) line corresponds to the approximated solution discovered in \cite{Scherrer:2007pu} (equation (23) in \cite{Scherrer:2007pu}) and the dash-dotted (dark) line corresponds to the approximated solution, \eqref{eqfinal}, presented here (assuming $\chi=0.1$).

\section{Exponential potential}\label{sec:II}

In order to motivate the analysis for a general (arbitrary) potential, let us consider the simpler case of the exponential potential $V(\phi)=V_0 e^{-\lambda \phi}$ \cite{Copeland:1997et,Halliwell:1986ja,Barreiro:1999zs,Burd:1988ss,Sami:2002fs,Rubano:2001su,Heard:2002dr,Coley:1997nk,Liddle:1988tb,Rubano:2003et,Guo:2003eu,Aguirregabiria:1993pm,Aguirregabiria:1993pk,Ibanez:1995zs,Pavluchenko:2003ge,Goheer:2002ac,Piedipalumbo:2011bj,Fang:2006zq,Aguirregabiria:1996uh,Copeland:2009be}.

In order to do the analysis from the dynamical systems viewpoint of the mixture of a scalar field with exponential potential and a Chaplygin gas, we need to consider the variables $x,y,s$ defined in the previous section  plus the new variable
$$z=\frac{A}{3H^2 \rho_{ch}^\alpha}.$$
Then, from the equations of
motions (\ref{ecH}),(\ref{ecch}) and (\ref{ecfi}) we obtain the
following autonomous system
\begin{align}
&x'=-3 x +\sqrt{\frac{3}{2}}\lambda y^2 +\frac{3}{2}x\left[1+x^2-y^2\right]-\frac{3}{2}x z,\nonumber\\
&y'=-\sqrt{\frac{3}{2}}\lambda x y+\frac{3}{2}y\left[1+x^2-y^2\right]-\frac{3}{2}y z,\nonumber\\
&z'=3z\left[1+\alpha+x^2-y^2\right]-3 z^2-\frac{3\alpha z^2}{1-x^2-y^2}. \label{autonomousab}
\end{align}
We note, that in the dust limit ($A\rightarrow 0$) the variable $z$ becomes automatically zero, the last equation in the system  \eqref{autonomousab}  is satisfied identically ($z=0$ defines an invariant set) and the remaining equations \eqref{autonomousab} corresponds to the usual exponential quintessence scenario \cite{Copeland:1997et}.

Now, it is convenient  to express the observable magnitudes in terms of the phase space variables. These observable magnitudes are the dimensionless Dark Energy density,  $\Omega_\phi,$ given by \eqref{Omegaphi}; the equation of state (EoS) parameter of the dark energy given by
\begin{equation}\label{wphi}
\omega\equiv\frac{P_\phi}{\rho_\phi}=\frac{x^2-y^2}{x^2+y^2};
\end{equation}
the EoS of the Chaplygin gas
\begin{equation}\label{wch}
\omega_{ch}\equiv\frac{P_{ch}}{\rho_{ch}}=-\frac{z}{1-x^2-y^2};
\end{equation}
the total (effective) EoS given by
\begin{equation}\label{wtot}
\omega_{tot}\equiv\frac{P_{tot}}{\rho_{tot}}=x^2-y^2-z;
\end{equation}
and the deceleration parameter
\begin{equation}\label{decc}
q\equiv -\frac{a\ddot a}{(\dot a)^2}=-1+\frac{3}{2}\left[1+x^2-y^2-z\right].
\end{equation}
These expressions are valid not only at the fixed points but also they are valid  in the whole phase space.
Thus,  we evaluate them at the fixed points in order to determine the type of solution that they represent.

 \begin{table*}[t!]
\begin{center}
\begin{tabular}{|c|c|c|c|c|c|c|c|c|c|}
\hline
&&&&& &&&&  \\
 Cr. P.& $x_c$ & $y_c$ & $z_c$  & Existence & $\Omega_{\phi}$ &  $\omega_{\phi}$ & $\omega_{ch}$ & $\omega_{tot}$ & $q$\\
\hline \hline
$A$& 0 & 0 & 0  & always
 & 0 & arbitrary & 0 & 0 & $\frac{1}{2}$ \\
\hline
 $B$& 1& 0 & 0 &  always  &
1  &   1 & arbitrary & 1 & 2\\
\hline
 $C$& -1& 0 & 0 &     always &
1  &   1 & arbitrary & 1 & 2\\
\hline
$D$& $\frac{\lambda}{\sqrt{6}}$ & $\sqrt{1-\frac{\lambda^2}{6}}$ &
0
 &  ${\lambda}^2\leq6$  &   1 &
$-1+\frac{\lambda^2}{3}$ & arbitrary & $-1+\frac{\lambda^2}{3}$ &
$-1+\frac{\lambda^2}{2}$\\
\hline
 $E$& $\sqrt{\frac{3}{2}}\frac{1}{\lambda}$ &
$\sqrt{\frac{3}{2(\lambda)^2}}$ &
0  &    $\lambda^2\geq3$ & $\frac{3}{\lambda^2}$ & 0 & 0 & 0 &
$\frac{1}{2}$\\
\hline
$F$&0 & 1 & 0&    $\lambda=0$  & 1 & $-1$ & arbitrary & $-1$ & $-1$\\
\hline
$G$& 0 & $y_c$ & $1-y_c^2$  &  $y_c^2<1, y_c\neq 0, \lambda=0$ & $y_c^2$ & $-1$ & $-1$ & $-1$ & $-1$\\
\hline
$K$ & 0 & 0 & 1 &    always  & 0 & arbitrary & $-1$ & $-1$ & $-1$\\ \hline
\end{tabular}
\end{center}
\caption[crit]{\label{critbexp} \it{The real and physically meaningful critical points of the autonomous system (\ref{autonomousab}) associated to the exponential potential. Stability conditions and the values of the dark-energy density parameter $\Omega_{DE}$, of the  dark-energy EoS parameter $\omega_{DE}$, of the total EoS parameter   $\omega_{tot}$ and of the deceleration parameter $q$.}  }
\end{table*}

 \begin{table*}[t!]
\begin{center}
\begin{tabular}{|c|c|c|}
\hline
&&   \\
 Cr. P.& Eigenvalues &
Stability \\
\hline \hline
$A$& $-\frac{3}{2},\, \frac{3}{2},\, 3(1+\alpha)$& saddle\\
\hline
 $B$& $3,\, 3-\sqrt{\frac{3}{2}}\lambda,\, \text{undef.}$ &    unstable \\
\hline
 $C$& $3,\, 3+\sqrt{\frac{3}{2}}\lambda,\, \text{undef.}$ &     unstable \\
\hline
$D$ &  $-3+\lambda^2,-\frac{1}{2}(6-\lambda^2),\, \text{undef.}$ & saddle  \\
\hline
 $E$& $3(1+\alpha), \beta^-, \beta^+$ & saddle   \\
\hline
$F$& $-3, -3, \text{undef.}$ & stable (see Appendix \ref{A.1})\\
\hline
$G$& $0,-3(1+\alpha), -3$ & NH \\
\hline
$K$& $-3,0, -3(1+\alpha)$ & stable\\
\hline
\end{tabular}
\end{center}
\caption[crit]{\label{crit2exp} \it{The real and physically meaningful
critical points of the autonomous system (\ref{autonomousab}) associated to the exponential potential.  Stability
conditions, NH stands for non-hyperbolic. Observe that the critical points $B, C, D$ and $F$ belong to the singular surface $x^2+y^2=1.$ In this case both denominator and numerator of (\ref{autonomousab}) vanish simultaneously. In this case the additional eigenvalue due to the extra $z$-coordinate could be finite positive or infinite with undefined sign depending of how the point is approached.  Thus, linear approximation fails and we need to resort to numerical investigation.}}
\end{table*}

In table \eqref{critbexp} we show  the existence
conditions for the real and physically meaningful
(curves of) critical points of the autonomous system (\ref{autonomousab}) associated to the exponential potential and also the values of the dark-energy
density parameter $\Omega_{DE}$, of the  dark-energy EoS parameter $\omega_{DE}$, of the total EoS parameter  $\omega_{tot}$ and of the deceleration parameter $q$ evaluated at them.

Now, let us discuss in more details the stability conditions for the corresponding critical points.
The critical point $A$ associated to a matter dominated universe is a saddle point.
Observe that the singular points $B, C, D$ and $F$ belong to the singular surface $x^2+y^2=1.$ In this case, both, the denominator and the numerator of Eq. (\ref{autonomousab}) vanish simultaneously. In this case the additional eigenvalue due to the extra $z$-coordinate could be finite positive or infinite with undefined sign depending of how the point is approached.  Thus, linear approximation fails and we need to resort to numerical investigation.
The critical points $B$ and $C,$ corresponding to stiff solutions, are always unstable. $B$ ($C$ resp.) is a local source for $\lambda>-\sqrt{6}$ ($\lambda<\sqrt{6}$, resp.), otherwise  they are saddles. This argument is supported by numerical studies as shown in Figures \ref{fig2} (a)-(d), for the values of the parameters in the typical intervals (that are determined by the bifurcation values).
Critical points $D$ and $E$ are the usual quintessence solutions widely investigated in the literature (see for instance \cite{Copeland:1997et}). Then, the main difference with respect to the results found in Ref. \cite{Copeland:1997et} is that, for $\alpha>0$ none of them can be a late time attractor. For $D$ this argument is based on numerical analysis, since $D$ belongs to the singular surface $x^2+y^2$ and in  this case, both the numerator and the denominator of the the system (\ref{autonomousab}) vanish, and then, the linear approximation is not valid.  In the case of the point $E$ we support this result in the fact that there exist one eigenvalues with positive real part. This means that if we include a GCG in the background, we cannot  get a stable solution dominated by the scalar field ($D$) or an scaling solution ($E$). If we restrict our attention to the invariant set $z=0$, then we have that $D$  is a stable one for $\lambda^2<3$  and thus it can be the late time state of the universe. In this case the equation for $z$ is vanished identically, and we do not require to include this variable  in the analysis. $D$ corresponds to a dark-energy dominated universe, with a dark-energy
EoS in the quintessence regime, which can be accelerating or not according to the $\lambda$-value. Additionally, this
solution is free of instabilities. This point is quite important, since it is stable and possesses $\omega_{DE}$ and $q$ compatible with
observations \cite{Copeland:1997et}. Point $E$ is stable in the invariant set $z=0.$ It can attract the universe at late times (in case of a GCG behaving as dust), and it is free of instabilities. It has the advantage that the dark-energy density parameter lies in the interval $0<\Omega_{DE}<1$, that is it can alleviate the coincidence problem, but it has the disadvantage that it is not accelerating and possesses $\omega_{DE}=0$, which are not favored by observations  \cite{Copeland:1997et}. However, let us remark that they are saddles for the full vector field. The solution $F$ exists for $\lambda=0.$ It represents a de Sitter solution which is stable but not asymptotically stable (see the Appendix \ref{A.1}). The curve of critical points $G$ (which exists only for $\lambda=0$) is stable but not asymptotically stable, whereas,  $K$ is asymptotically stable (see details of the center manifold calculations for both $G$ and $K$ in the Appendix \ref{appendixA}). To finish this section let us proceed to the discussion of some numerical elaborations:

\begin{figure*}[t!]
\begin{center}
\begin{tabular}{cc}
\includegraphics[scale=0.5]{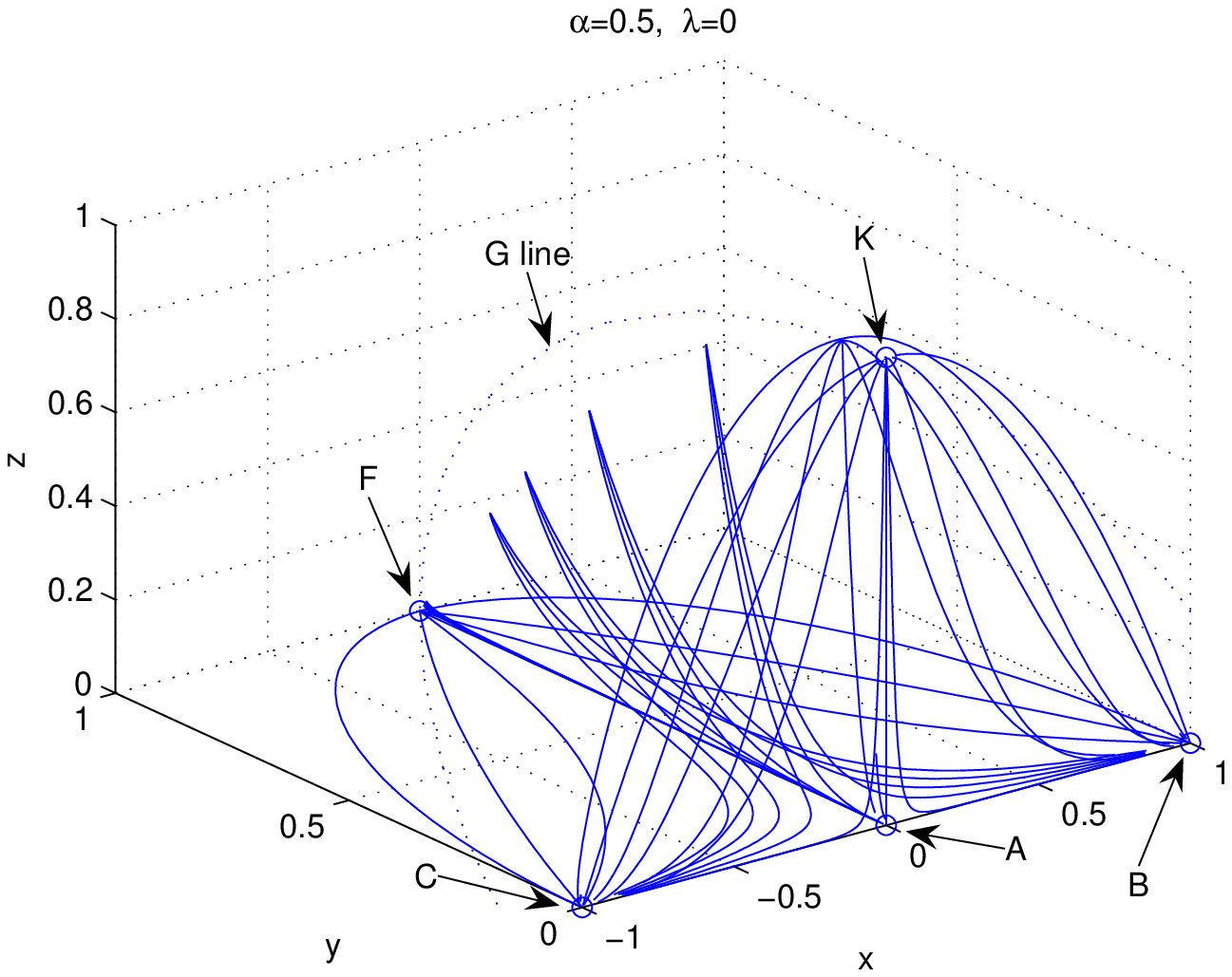} & \includegraphics[scale=0.5]{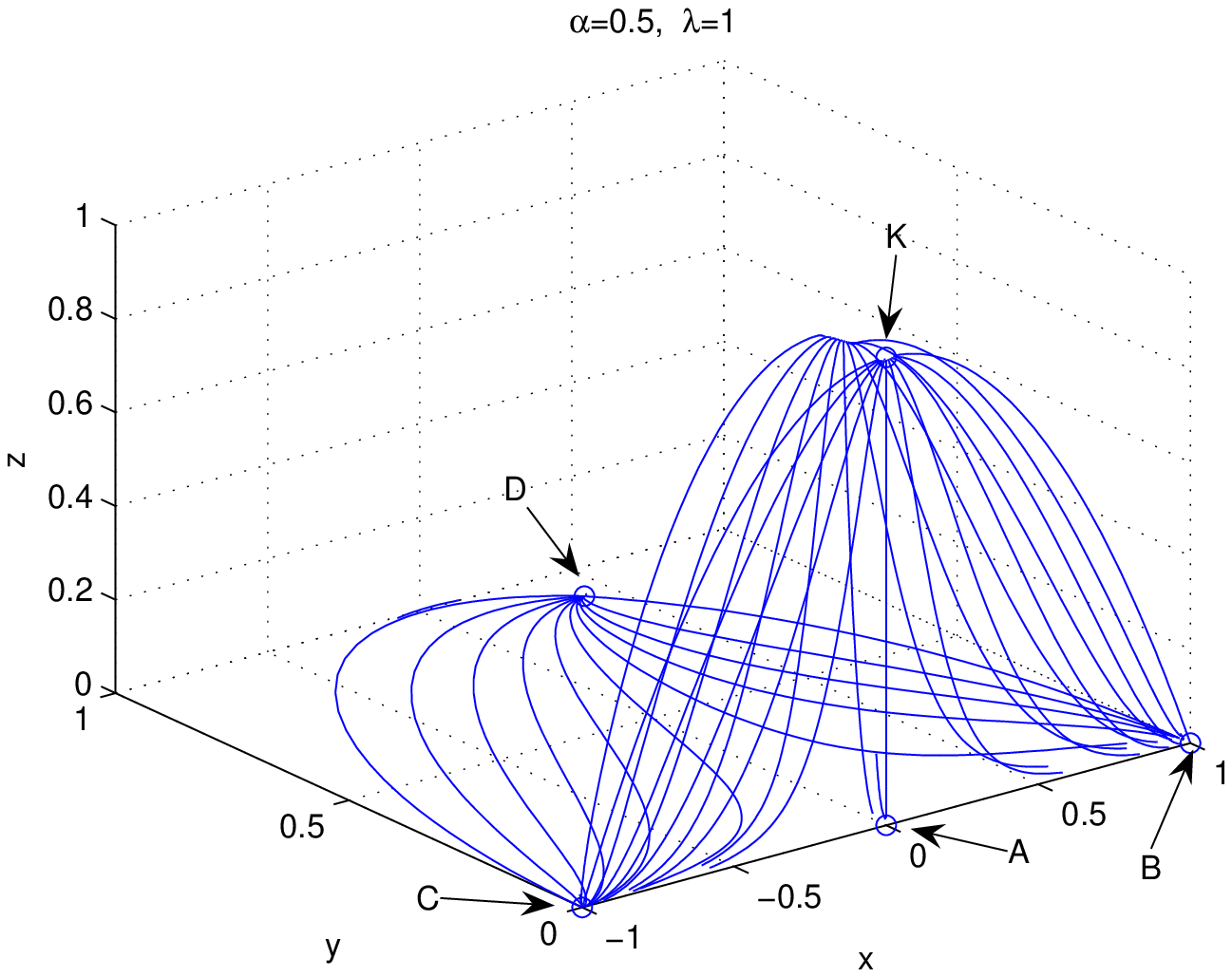} \\
(a) & (b) \\
\includegraphics[scale=0.5]{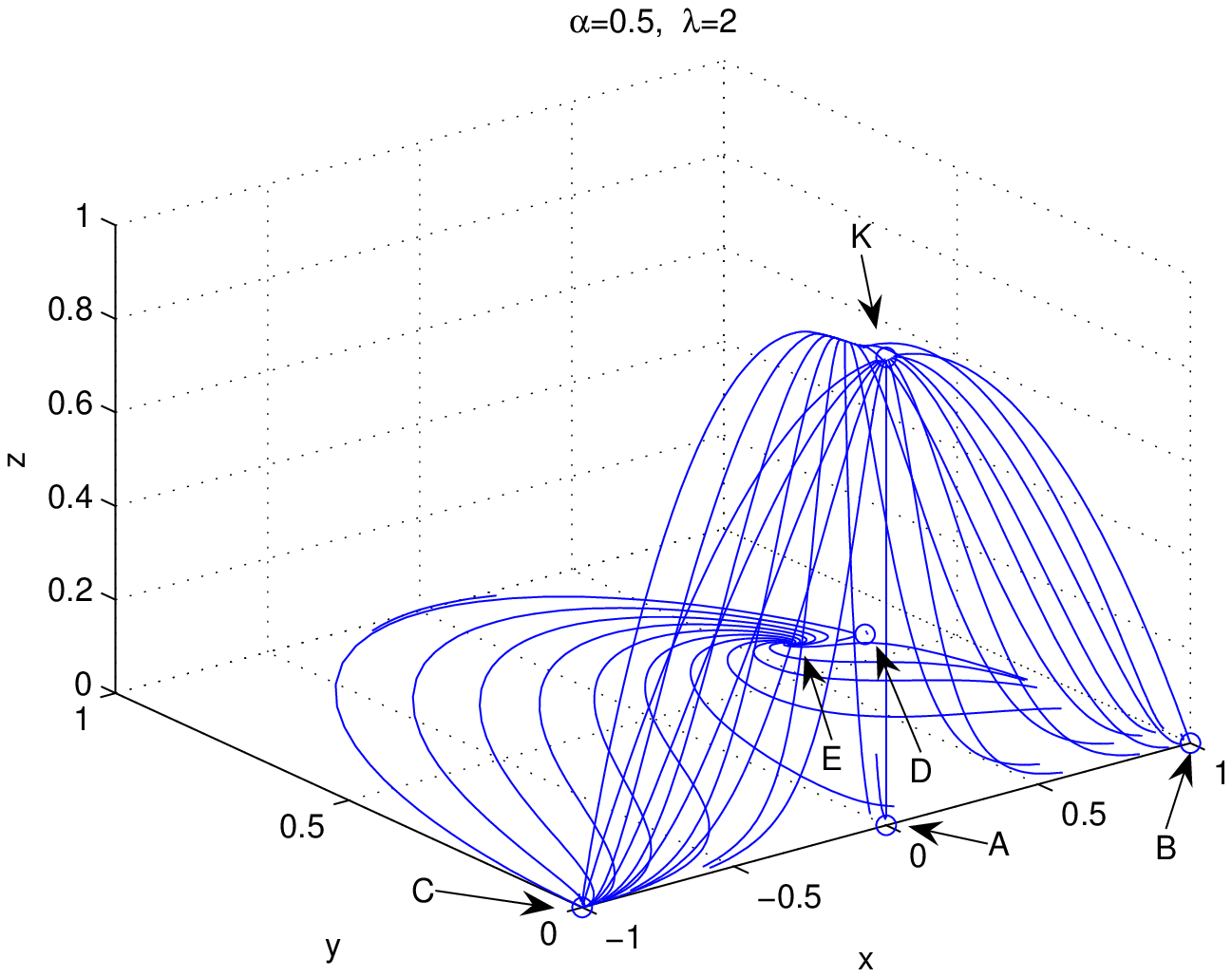} & \includegraphics[scale=0.5]{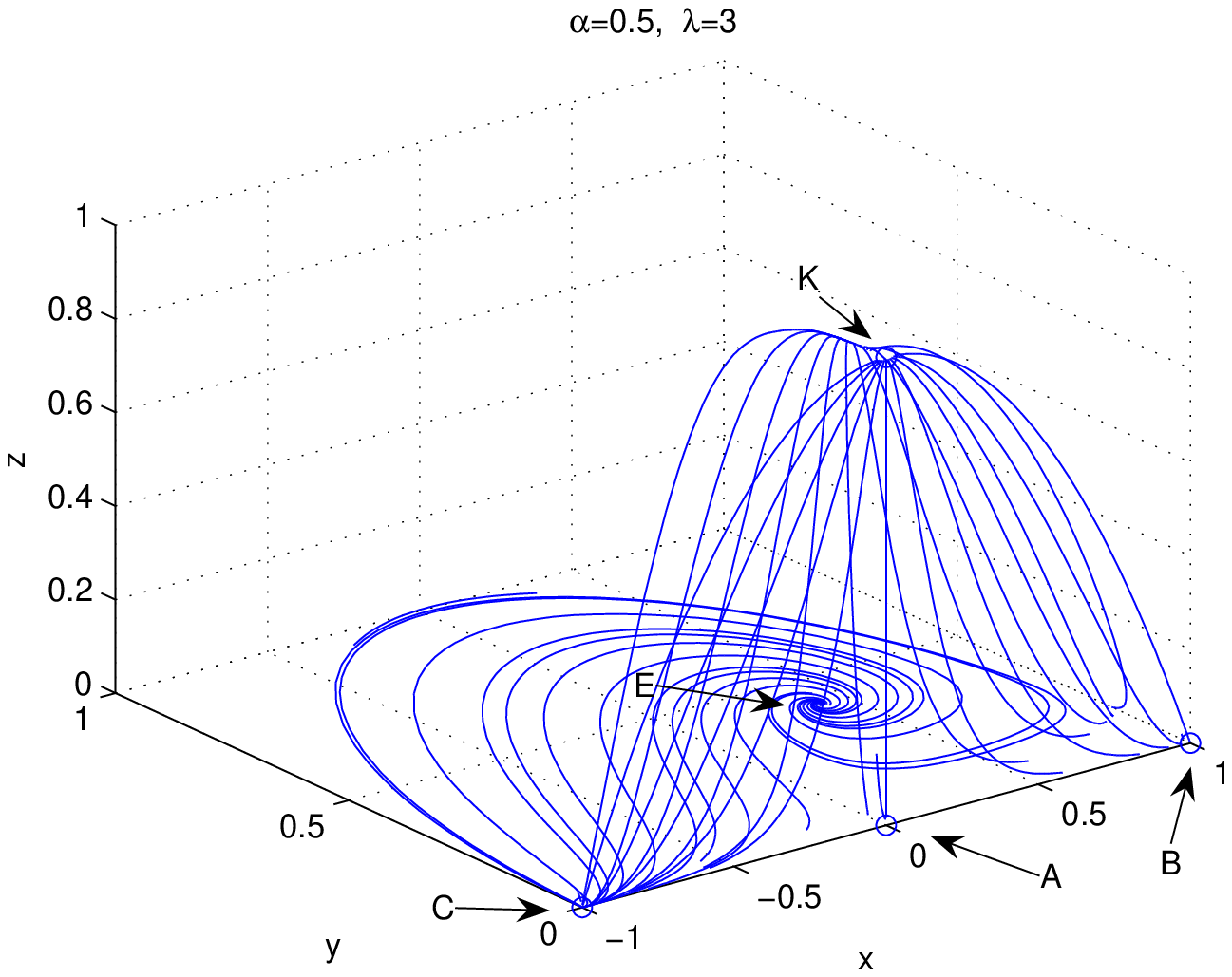}\\
(c) & (d)
\end{tabular}
\caption{\label{fig2} \it{The phase space of the system  (\ref{autonomousab}). Without lack of generality we use $\alpha=0.5$}:\\
(a) \it{$\lambda=0.$  $B$ and $C$ are local sources; $A$ is a saddle.  $F$ coincides with $D$ (contained in the curve $G$) and it is stable but not asymptotically stable for $f(0)\geq 0,$  otherwise it is a saddle (see Appendix \ref{A.2}). $E$ does not exist. Any arbitrary point in the curve $G$  (that exist only for $\lambda=0$) is stable, attracting an open set of orbits. The center manifold of $K$ is stable, also $K$ is.}\\
(b) {\it{ $\lambda=1.$ $B$ and $C$ are local sources; $A$ and $D$ are saddles and $K$ is the attractor. $D$ is a local attractor in the invariant set $z=0.$ The solutions at the $x$-$y$ plane correspond to those at Figure 2 in \cite{Copeland:1997et}}.}\\
(c) {\it{$\lambda=2.$ $B$ and $C$ are local sources; $A,$ $D$ and $E$ are saddles and $K$ is the attractor. $E$ is a local attractor in the invariant set $z=0.$ The solutions at the $x$-$y$ plane correspond to those at Figure 3 in \cite{Copeland:1997et}.}}\\
(d) {\it{$\lambda=3.$ $B$ (the kinetic-dominated solution with $\lambda x>0$) is a saddle point; $C$  (the kinetic-dominated solution with $\lambda x<0$) is the local source (unstable node); $D$ does not exists; $A$ and $E$ are saddles and $K$ is the attractor. $E$ is a local (spiral) attractor  in the invariant set $z=0.$ The solutions at the $x$-$y$ plane corresponds to those at Figure 4 in \cite{Copeland:1997et}. }} }
\end{center}
\end{figure*}

\begin{itemize}

\item Fig. \ref{fig2} (a) shows several orbits for the values of the parameters $\alpha=0.5, \lambda=0.$  $B$ and $C$ are local sources; $A$ is a saddle.  $F$ coincides with $D$ (contained in the curve $G$) and it is stable but not asymptotically stable for $f(0)\geq 0,$  otherwise it is a saddle (see Appendix \ref{A.2}). $E$ does not exist. Any arbitrary point in the curve $G$  (that exist only for $\lambda=0$) is stable, attracting an open set of orbits. The center manifold of $K$ is stable.

\item In Fig. \ref{fig2} (b) we  presented several orbits in the phase space  for $\alpha=0.5, \lambda=1.$ The kinetic-dominated solution $B$ and $C$ are local sources; the matter dominated solution $A$ and the scalar-field dominated solution $D$ are saddles and the  Chaplygin gas dominated solutions (which also mimics a
de Sitter solution) $K$ is the attractor. $D$ is a local attractor in the invariant set $z=0.$

\item Fig. \ref{fig2} (c) shows several orbits in the phase space for $\alpha=0.5, \lambda=2.$ $B$ and $C$ are local sources; $A,$ $D$ and $E$ are saddles and $K$ is the attractor. $E$ is a local attractor in the invariant set $z=0.$

\item  Finally, in   Fig. \ref{fig2} (d) we display several orbits in the phase space for $\alpha=0.5, \lambda=3.$ $B$ (the kinetic-dominated solution with $\lambda x>0$) is a saddle point; $C$  (the kinetic-dominated solution with $\lambda x>0$) is the local source (unstable node); $D$ does not exists; $A$ and $E$ are saddles and $K$ is the attractor. $E$ is a local (spiral) attractor  in the invariant set $z=0.$

\end{itemize}

Now, as commented before, it is a fact that the dynamical evolution leave the coordinates $(x,y)$ within the upper-half unit disc. However, the only restriction on $z$ is that $z\geq 0.$ This means that in priciple the $z$-coordinate could be unbounded and, then,  there migh exist critical points at infinity (which would correspond to $z\rightarrow +\infty$).
In order to determine the fixed points at infinity and study their stability, we need to compactify the phase space using the
Poincar\'e method. Transforming to polar coordinates $(r(\tau), \theta(\tau), \psi(\tau)))$ \cite{Lefschetz,Carloni:2004kp,Abdelwahab:2007jp}: 
\begin{equation}
x=r \cos \theta \sin \psi, y= r \cos \theta \sin \psi, z= r \cos \psi, 
\end{equation}
where $0\leq\psi\leq\frac{\pi}{2}, 0\leq \theta \leq\pi$ and substitutig $r=\frac{R}{1-R},$ the regime $r \rightarrow \infty$ corresponds to $R \rightarrow 1$. 
The points $x,y,z$ are mapped onto 
\begin{equation}
x_R= R \cos \theta \sin \psi, y_R= R \cos \theta \sin \psi, z_R= R \cos \psi, 
\end{equation} thus, the points at infinity are mapped on the unitary sphere $R=1.$

Using this
coordinate transformation, introducing the new time variable  $d\eta=\frac{d\tau}{(1-R)},$ which preserves the time orientation, the leading terms of the system (\ref{autonomousab}) as $R\rightarrow 1$  are
\begin{align}
&R'\rightarrow\frac{3 \left(\cos (2 \theta ) (\cos (2 \psi )+3) \sin ^2\psi\right)}{4},\\
&\theta'\rightarrow -{\sqrt{\frac{3}{2}} \lambda  \sin \theta \sin \psi }, \\
&\psi'\rightarrow -\frac{3 \left(\cos (2 \theta ) \cos\psi \sin ^3\psi \right)}{2 (1-R)},
\end{align}	where now the comma denotes derivative with respect $\eta.$ In this case the
radial equation does not contain the radial coordinate, thus, the fixed points can be obtained
using just the angular equations. Setting $\theta' = 0$ and $\psi' = 0$ we obtain that the fixed point with physical sense ($0\leq \Omega_\phi\leq 1$) must satisfy $\psi=0,$ i.e., $(x_R,y_R,z_R)=(0,0,1).$ In this case the eigenvalues of the Jacobian matrix associated to the angular coordinates are $\{0, 0\},$  and $R' = 0$ at the equilibrium point. Then, we cannot obtain information on their stability using the linearization. The complete analysis is outside the scope of the present investigation.
 
\section{Phase-space analysis without potential
specification}
\label{newmethod}

In order to transform the system \eqref{eqx}-\eqref{eqs}  to an autonomous one, first, it is necessary to determine a specific potential form
$V(\phi)$ of  the scalar field $\phi$. However, using the above example as a motivation, one could alternatively handle the potential differentiations using
the auxiliary
variable $s$ given by \begin{equation}\label{sdef}
s=-\frac{V'(\phi )}{V(\phi
)},\end{equation}  while keeping the potential still arbitrary \footnote{The variable $s$ is just a constant ($s\equiv\lambda$) for the exponential potential $V=V_0 e^{-\lambda \phi}$.}. The next step is to introduce the function  \begin{equation}
f\equiv s^2\,(\Gamma-1)=\frac{V''(\phi
   )}{V(\phi )}-\frac{V'(\phi )^2}{V(\phi )^2},
\label{fdef}
\end{equation}
to be an arbitrary function of $s.$
In fact, if $f$ can be expressed as an explicit one-valued function of $s$,
that is $f=f(s)$,  then, it is possible to write a  closed dynamical
system for $s$  and a set of normalized-variables.
On the other hand, by giving $f(s),$ non identically equal to zero, we obtain the expressions \begin{eqnarray}
\phi(s)&=&\phi_0-\int_{s_0}^s \frac{1}{f(K)} \, dK, \label{quadphi}\\
V(s)&=&e^{\int_{s_0}^s \frac{K}{f(K)} \, dK} \bar{V}_0\label{quadV},
\end{eqnarray} where the integration constants satisfy $V(s_0)=\bar{V}_0$,
$\phi(s_0)=\phi_0$ \footnote{We would like to note that the requirement that $f$ must be different to zero exclude of this analysis the case of the exponential potential and for this reasons we studied the exponential potential in the previous section separately.}. Thus, it is possible to reconstruct the potential $V$ by the elimination of $s$  between \eqref{quadphi} and \eqref{quadV}.
For the usual cosmological cases the potential can be written explicitly, that is $V=V(\phi)$.
The details of the method, coined ``Method of $f$-devisers'', were presented in \cite{Escobar:2013js}. This method has the significant advantage, that one can
first perform the analysis for arbitrary potentials and then just
substitute the desired forms, instead of repeating the whole procedure for
every distinct potential (see \cite{Escobar:2013js} and references therein).

Then, from the equations of
motions (\ref{ecH}), (\ref{ecch}) and (\ref{ecfi}) we result in the
following autonomous system
\begin{align}
&x'=-3 x +\sqrt{\frac{3}{2}}s y^2 +\frac{3}{2}x\left[1+x^2-y^2\right]-\frac{3}{2}x z,\nonumber\\
&y'=-\sqrt{\frac{3}{2}}s x y+\frac{3}{2}y\left[1+x^2-y^2\right]-\frac{3}{2}y z,\nonumber\\
&z'=3z\left[1+\alpha+x^2-y^2\right]-3 z^2-\frac{3\alpha z^2}{1-x^2-y^2},\nonumber\\
&s'=-\sqrt{6}f(s) x. \label{autonomousb}
\end{align}

 \begin{table*}[ht]
\begin{center}
\begin{tabular}{|c|c|c|c|c|c|c|c|c|c|c|}
\hline
&&&&& &&&&&  \\
 Cr. P.& $x_c$ & $y_c$ & $z_c$ & $s_c$ & Existence & $\Omega_{\phi}$ &  $\omega_{\phi}$ & $\omega_{ch}$ & $\omega_{tot}$ & $q$\\
\hline \hline
$A$& 0 & 0 & 0  & $s_c$ &  always
 & 0 & arbitrary & 0 & 0 & $\frac{1}{2}$ \\
\hline
 $B(s^*)$& 1& 0 & 0 & $s^*$ & always  &
1  &   1 & arbitrary & 1 & 2\\
\hline
 $C(s^*)$& -1& 0 & 0 &  $s^*$ &   always &
1  &   1 & arbitrary & 1 & 2\\
\hline
$D(s^*)$& $\frac{s^*}{\sqrt{6}}$ & $\sqrt{1-\frac{{(s^*)}^2}{6}}$ &
0
& $s^*$ &  ${(s^*)}^2\leq6$  &   1 &
$-1+\frac{(s^*)^2}{3}$ & arbitrary & $-1+\frac{(s^*)^2}{3}$ &
$-1+\frac{(s^*)^2}{2}$\\
\hline
 $E(s^*)$& $\sqrt{\frac{3}{2}}\frac{1}{s^*}$ &
$\sqrt{\frac{3}{2(s^*)^2}}$ &
0 &  $s^*$ &    $(s^*)^2\geq3$ & $\frac{3}{(s^*)^2}$ & 0 & 0 & 0 &
$\frac{1}{2}$\\
\hline
$F$&0 & 1 & 0& 0 &   always  & 1 & $-1$ & arbitrary & $-1$ & $-1$\\
\hline
$G$& 0 & $y_c$ & $1-y_c^2$ & 0 &  $y_c^2<1, y_c\neq 0$ & $y_c^2$ & $-1$ & $-1$ & $-1$ & $-1$\\
\hline
$K$ & 0 & 0 & 1& $s_c$ &    always  & 0 & arbitrary & $-1$ & $-1$ & $-1$\\ \hline
\end{tabular}
\end{center}
\caption[crit]{\label{critb} \it{The real and physically meaningful
(curves of) critical points of the autonomous system (\ref{autonomousb}). Existence
conditions and the values of the dark-energy
density parameter $\Omega_{DE}$, of the  dark-energy EoS
parameter $\omega_{DE}$, of the total EoS parameter
  $\omega_{tot}$ and of the deceleration parameter $q$. We use the notation $s^*$ for the values of $s=s^*$ such that $f(s^*)=0,$ and $s_c$ for denoting arbitrary values of $s$ at equilibrium.} }
\end{table*}

 \begin{table*}[ht]
\begin{center}
\begin{tabular}{|c|c|c|}
\hline
&&   \\
 Cr. P.& Eigenvalues &
Stability \\
\hline \hline
$A$& $-\frac{3}{2},\, \frac{3}{2},\, 0,\, 3(1+\alpha)$& saddle\\
\hline
 $B(s^*)$& $3,\, 3-\sqrt{\frac{3}{2}}s^*,\, -\sqrt{6}f'(s^*),\, \text{undef.}$ &     unstable \\
\hline
 $C(s^*)$& $3,\, 3+\sqrt{\frac{3}{2}}s^*,\, \sqrt{6}f'(s^*),\, \text{undef.}$ &     unstable \\
\hline
$D(s^*)$ &  $-3+(s^*)^2,-\frac{1}{2}(6-(s^*)^2),-s^*f'(s^*),\, \text{undef.}$ & saddle \\
\hline
 $E(s^*)$& $3(1+\alpha), \beta^-(s^*), \beta^+(s^*),  -\frac{3 f'(s^*)}{s^*} $ & saddle \\
\hline
$F$& $\text{undef.}, \text{undef.}, \delta^+, \delta^-$ & stable (see Appendix \ref{A.2})\\
\hline
$G$& $0,-3(1+\alpha), -3, \Delta^+, \Delta^-$ & NH, stable for $f(0)>0, y_c>0$, saddle otherwise  \\
\hline
$K$& $-3,0, 0, -3(1+\alpha)$ & NH (unstable)\\
\hline
\end{tabular}
\end{center}
\caption[crit]{\label{crit2} \it{The real and physically meaningful
critical points of the autonomous system (\ref{autonomousb}). Stability
conditions, NH stands for non-hyperbolic. We introduce the notations $\beta^\pm(s^*)=\frac{3}{4} \left(-1\pm\frac{\sqrt{24
(s^*)^2-7 (s^*)^4}}{(s^*)^2}\right),$ $\delta^\pm=-\frac{3}{2}\left(1\pm\sqrt{1-\frac{4}{3}f(0)}\right),$ and $\Delta^\pm=-\frac{3}{2}\left(1\pm\sqrt{1-\frac{4}{3} y_c^2 f(0)}\right).$ Observe that the critical points $B(s^*), C(s^*), D(s^*)$ and $F$ belong to the singular surface $x^2+y^2=1.$ In this case both denominator and numerator of (\ref{autonomousb}) are vanished simultaneously. In this case the additional eigenvalue due to the extra $z$-coordinate could be finite positive or infinite with undefined sign depending of how the point is approached.   For $F$ there are two eigenvalues whose nature depends on the  way that $F$ is approached. For that reason they are undefined. Thus, linear approximation fails and we need to resort to numerical works.}}
\end{table*}

In  table \ref{critb} we present the existence conditions for the real and physically meaningful (curves of) critical points of the autonomous system (\ref{autonomousb}). We use the notation $s^*$ for the values of $s=s^*$ such that $f(s^*)=0,$ and $s_c$ for denoting arbitrary values of $s$ at equilibrium.  We display also the corresponding values of the dark-energy density parameter $\Omega_{\phi}$, of the  dark-energy EoS parameter $\omega_{\phi}$, of the EoS of Chaplygin gas $\omega_{ch}$, of the total EoS   $\omega_{tot}$ and of the deceleration parameter $q$. In table \ref{crit2} are presented the stability conditions for the the critical points.

Now, let us comment briefly on the stability and physical interpretation of the critical points of \eqref{autonomousb}.

The curve of critical point $A$ is always a saddle. It represents cosmological solutions dominated by the Chaplygin gas mimicking dust, this solution correlates with the transient matter dominated epoch of the universe.
Observe that the critical points $B(s^*), C(s^*), D(s^*)$ with  $s^*$  such that $f(s^*)=0$ and $F$ belong to the singular surface $x^2+y^2=1.$ In this case both denominator and numerator of (\ref{autonomousb}) are vanished simultaneously. In this case the additional eigenvalue due to the extra $z$-coordinate could be finite positive or infinite with undefined sign depending of how the point is approached.  For $F$ there are two eigenvalues whose nature depends on the  way that $F$ is approached, for that reason they are undefined.

For $s^*$, the solutions $B(s^*)$ and $C(s^*)$ are past attractors or saddle points under the same conditions of the standard quintessence scenario \cite{Copeland:1997et} with the identification $s^*\equiv \lambda$ (see table \ref{crit2}). They represent solutions dominated by the kinetic energy of the scalar field mimicking a stiff fluid. The solutions $D(s^*), E(s^*)$ and $F$ represents the scalar field dominated solution, the matter-scalar scaling solution and de Sitter solutions dominated by the potential energy of the scalar field,  respectively. The main difference here with respect the standard quintessence scenario \cite{Copeland:1997et}  is that  $D(s^*), E(s^*)$ are saddle points (we are considering $\alpha>0$). So, the standard quintessence solutions are not late time solutions in this scenario.  For analyzing the important critical point $F$, the linear approximation fails and we need to resort to numerical studies or include higher order terms in the analysis. In fact,  following our approach in the Appendix \ref{A.2}, we find that actually $F$ is the late-time attractor for $f(0)>0$ and a saddle for $f(0)<0.$

Combining expressions \eqref{omegach1} and \eqref{wch} we find that as $\tau\rightarrow +\infty,$ $z\rightarrow 1-x^2-y^2.$ Thus at late times we can approximate the system \eqref{autonomousb} by
\begin{align}\label{aproxxys}
& x'=-3x+3x^3 +\sqrt{\frac{3}{2}}s y^2,\nonumber\\
& y'=-\sqrt{\frac{3}{2}}s xy+3 y x^2,\nonumber\\
& s'=-\sqrt{6} f(s) x,
\end{align}
and the decoupled equation
\begin{equation}\label{approxz}
z'=6 z x^2.
\end{equation}
If $x\rightarrow x_c\neq 0$ as $\tau\rightarrow +\infty,$ then from equation \eqref{approxz} follows that $z$ increases without bound in contradiction with the boundedness of $1-x^2-y^2.$ Thus, as time goes forward, $x\rightarrow 0.$ Hence $z\rightarrow 1-y_c^2$ where $0\leq y_c\leq 1.$
By calculating the critical points of the system \eqref{aproxxys} and analyzing their linear stability we find that the only candidates to be the late-time attractors are:
\begin{itemize} \item the curve $G$ which have the following system of eigenvalues and eigenvectors:
$$
\left(\begin{array}{ccc}
 0, & \Delta^+, & \Delta^- \\
 \{0,1,0\}, & \left\{-\frac{\Delta^+}{\sqrt{6} f(0)},0,1\right\}, & \left\{-\frac{\Delta^-}{\sqrt{6} f(0)},0,1\right\}
\end{array}\right),
$$ where $\Delta^\pm=-\frac{3}{2}\left(1\pm\sqrt{1-\frac{4}{3} y_c^2 f(0)}\right).$
Since the center subspace is tangent to the $y$-axis, follows that the curve is normally hyperbolic \footnote{Recall that a set of non-isolated critical points is said to be normally
hyperbolic if the only eigenvalues with zero real parts are those
whose corresponding eigenvectors are tangent to the set. In this case the stability of the set can be deduced by examining the signs of the remaining non-null eigenvalues (i.e., for a curve, in the remaining $n-1$
directions) \cite{Aulbach1984a}.}. Then follows the stability of $G$ on the space $(x,y,s).$ This argument is not complete, since we have forget about what happens in the $z$-direction. In fact, in the general case (when the $z$-direction is included in the analysis), this curve is actually non-hyperbolic and it is not normally hyperbolic anymore, thus we cannot obtain information about its stability looking at the linearization. This one is the main difference that appears when considering the extra direction $z$.
\item The other candidate is the curve $K$ which have the following system of eigenvalues and eigenvectors $$
\left(\begin{array}{ccc}
 -3, & 0, & 0 \\
 \left\{\frac{\sqrt{\frac{3}{2}}}{f(s_c)},0,1\right\}, & \{0,0,1\}, & \{0,1,0\}
\end{array}\right), $$
which is also normally hyperbolic (the center subspace is the plane $y$-$s$ is tangent to the line $s=s_c$).

\item Finally, both numerical simulations and analytical methods suggest that $F$ (contained in the curve $G$) is an attractor for $f(0)>0,$ and for  $f(0)<0,$  it is a saddle (see Appendix \ref{A.2}).
\end{itemize}
The above heuristic reasoning suggest that the future attractor of the system \eqref{autonomousb} is located at the curve $G$, which contains the especial point $F$, or it is located at the curve $K$.

With the exception of the point $F,$ which is dominated by a constant potential, $G$ represents a class of solutions where neither the potential energy of the scalar field nor the Chaplygin gas dominates. On the other hand the curve $K$ corresponds to purely Chaplygin gas dominated solutions (which also mimics a de Sitter solution).

Indeed, using the Center Manifold Theory it can be proved that if the condition $f(0)>0,$ is satisfied, the curve of fixed points $G$ is stable but not asymptotically stable. Applying the same procedure to the curve $K$ we find  that also the curve $K$ is stable but not asymptotically stable. The details of the calculation are presented in the Appendix \ref{appendixB}.

To investigate the dynamics at infinity one introduces the Poincar\'e variables \cite{Lefschetz,Carloni:2004kp,Abdelwahab:2007jp}:
\begin{align}
&x= \frac{R}{1-R} \cos \theta \sin \varphi \sin \psi,
y= \frac{R}{1-R} \sin \theta \sin \varphi \sin \psi,\nonumber\\
&z= \frac{R}{1-R} \sin \varphi  \cos \psi, 
s=\frac{R}{1-R} \cos \varphi,\label{poincarevars}
\end{align} where $0\leq\psi\leq\frac{\pi}{2}, 0\leq \theta \leq\pi$  and $0\leq \varphi\leq \pi,$  and the new time variable  $d\eta=\frac{d\tau}{(1-R)},$ which preserves the time orientation.
The region at infinity $x^2+y^2+z^2+s^2\rightarrow \infty$ corresponds to the region $R\rightarrow 1$  in the $R,\theta,\varphi,\psi$ space. 
Then, we take te limit $R\rightarrow 1$ in $R',\theta',\varphi',\psi',$ where now the comma denotes derivative with respect to $\eta,$  and preserve the leading terms. 
In the case that the radial equation does not contain the radial coordinate, the fixed points can
be obtained using just the angular equations. Setting $\theta '=0$, $\psi '=0$ and $ \varphi '=0$, are
obtained the fixed points.
The stability of these points is studied by analyzing first the stability of the angular coordinates
and then deducing, from the sign of $R'$, the stability on the radial direction
\cite{Lefschetz,Carloni:2004kp,Abdelwahab:2007jp}. That is, it is required $R'>0$ at equilibrium. This means that the radial coordinate increases in value to reaching the boundary $R=1$ from below. To do the analysis it is required to provide the functional form of $f(s),$ however, the complete analysis is outside the scope of the present study. 

\subsection{An example: Cosh-like potential}

The cosh-like potential $V(\phi)=V_{0}\left[\cosh\left( \xi \phi
\right)-1\right]$ has been widely studied in the literature (see for example \cite{Sahni:1999qe,Sahni:1999gb,Lidsey:2001nj,Pavluchenko:2003ge,
Copeland:2009be,Ratra:1987rm,Wetterich:1987fm,Matos:2000ng,Sahni:1999qe,Wetterich:1987fm,Ratra:1987rm,Copeland:2009be,Matos:2009hf,Leyva:2009zz,Escobar:2013js}).
 For this potential,
\begin{equation}\label{fs}
f(s)=-\frac{1}{2}(s-\xi)(s+\xi).
 \end{equation}
 Observe that $f(0)=\frac{1}{2}\xi^2>0.$ This is the sufficient condition for the stability of the class of de Sitter solutions represented by the curve of critical points $G.$
For this choice $s^*\in\left\{\xi, -\xi\right\}.$ Also $f'(s)=-s,$ thus  $f'(\xi)=-\xi, f'(-\xi)=\xi.$
For this choice the system \eqref{autonomousb} admits twelve (curves of) critical points denoted by  $A, B(\xi), B(-\xi), C(\xi), C(-\xi), D(\xi), D(-\xi), E(\xi), E(-\xi)$, $F, G$ and $K.$

\begin{figure*}
\begin{center}
\begin{tabular}{cc}
\includegraphics[scale=0.5]{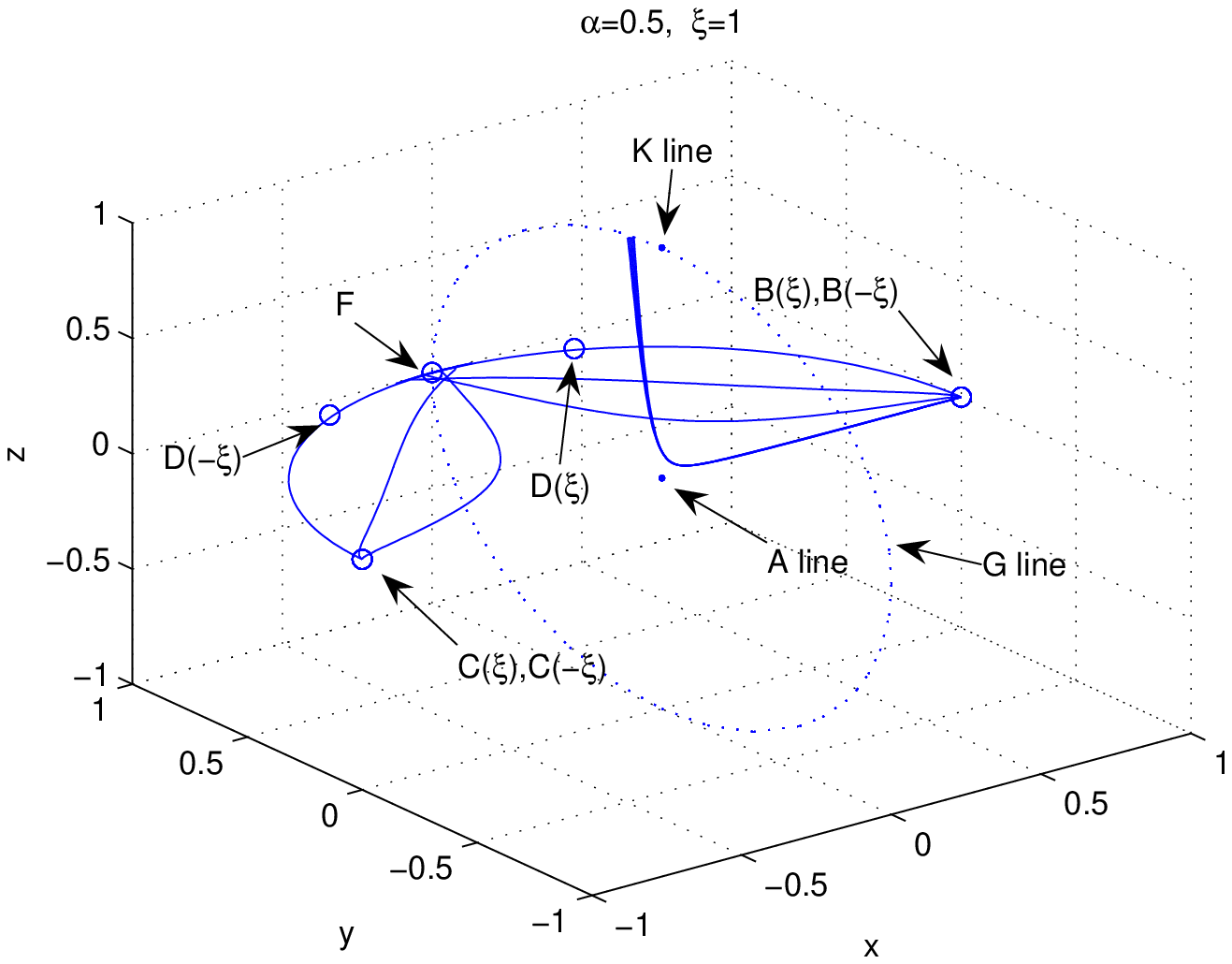} &
\includegraphics[scale=0.5]{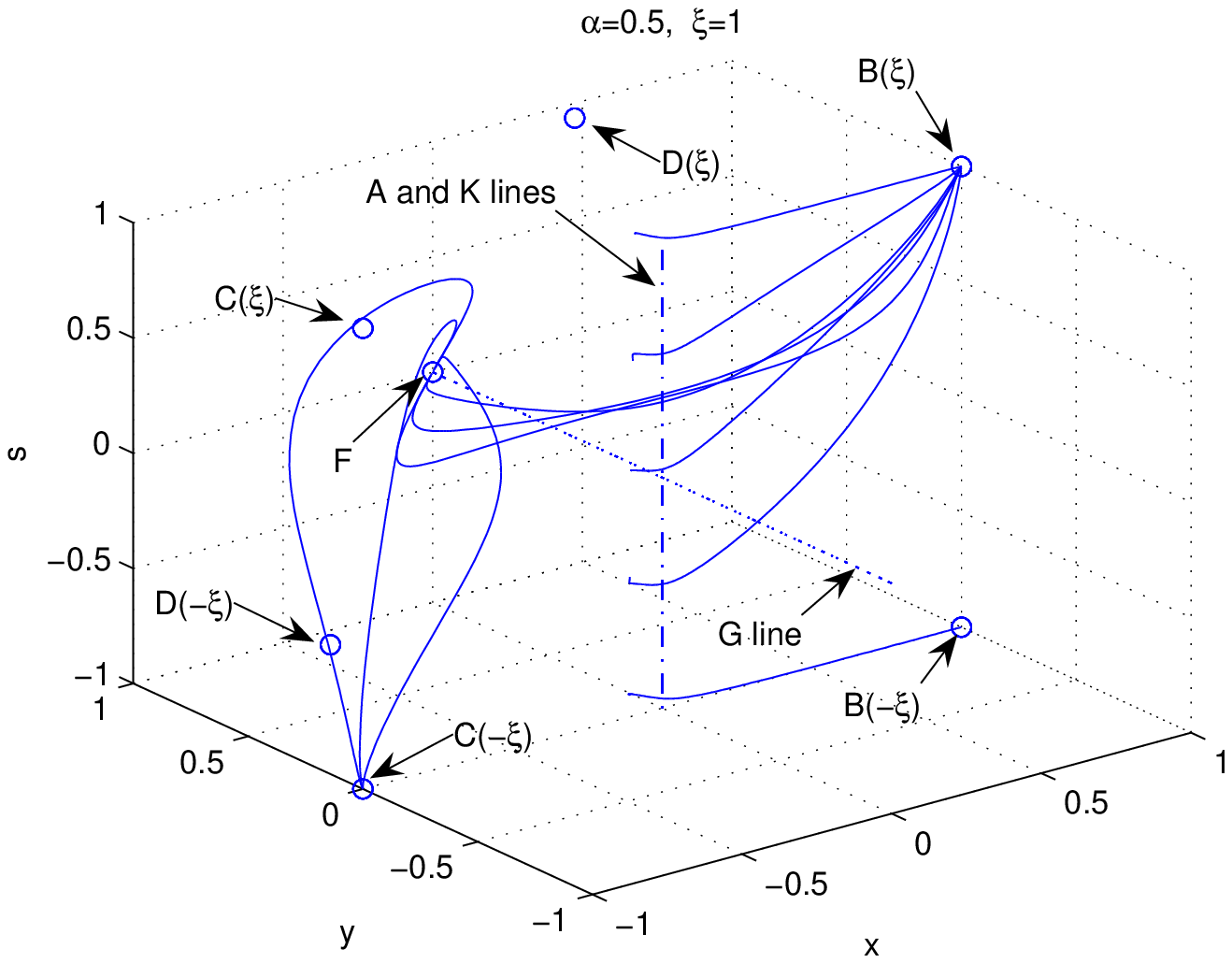} \\
\includegraphics[scale=0.5]{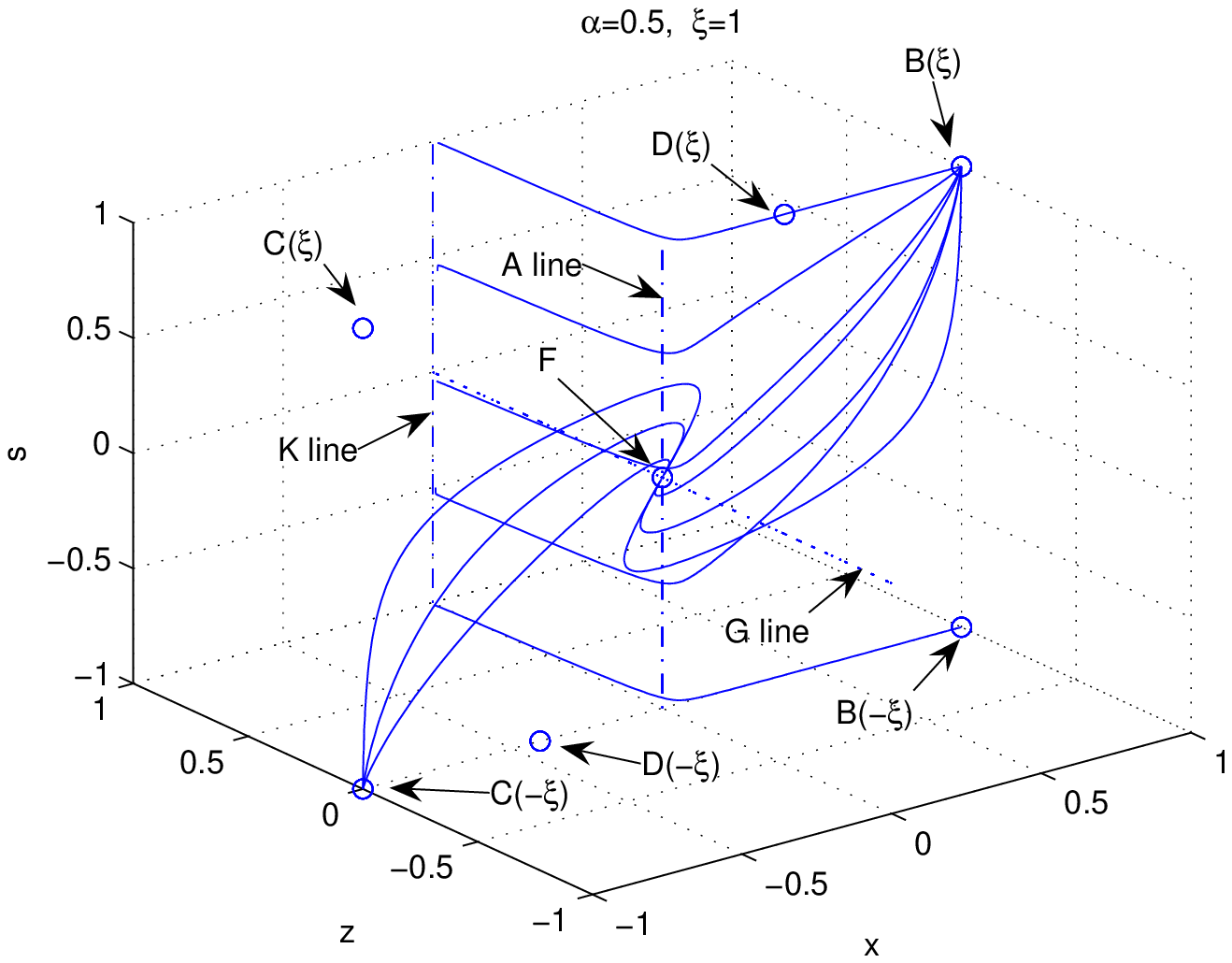} &
\includegraphics[scale=0.5]{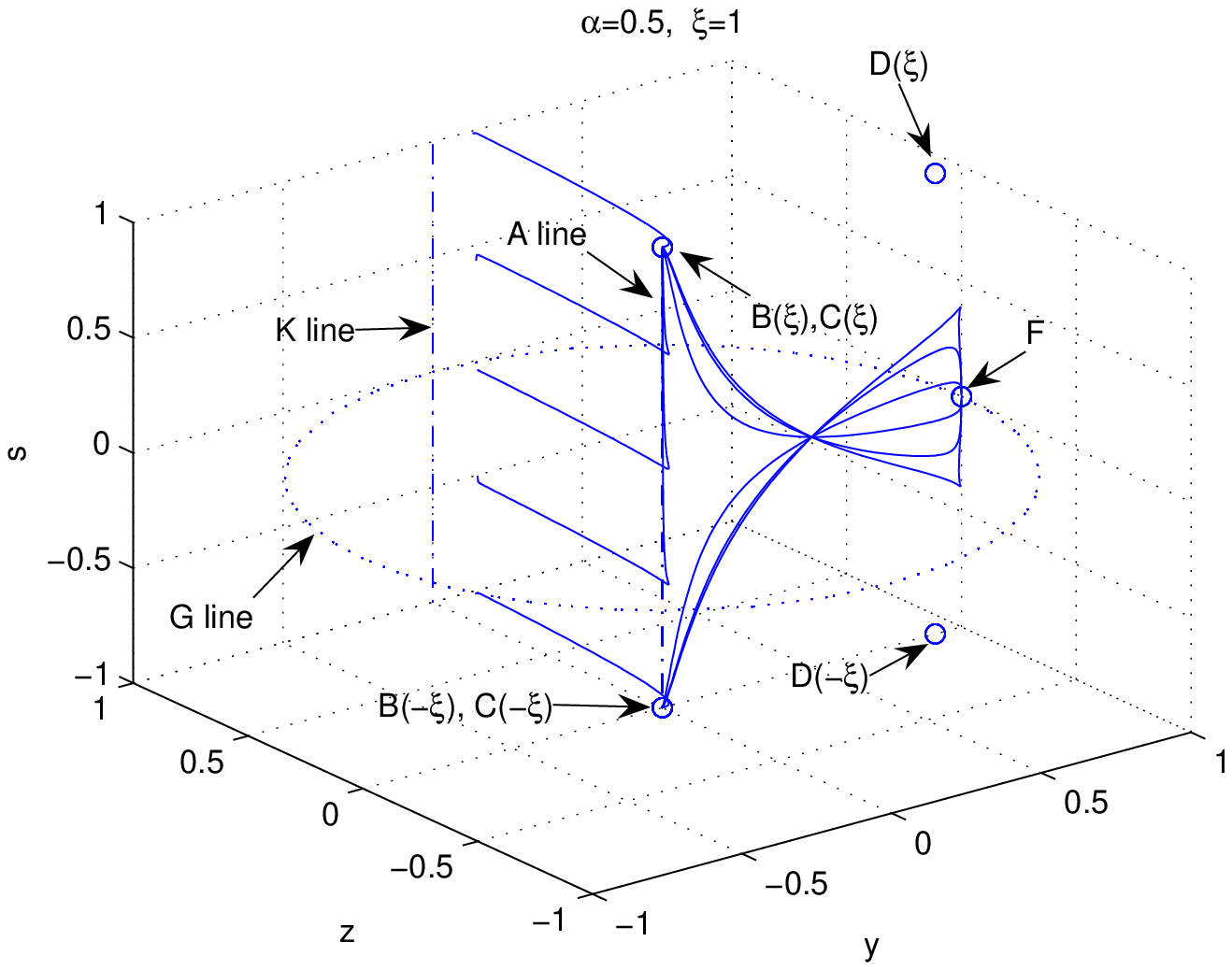}
\end{tabular}
\caption{\label{Fig4} \it{Projections of some orbits in the phase space of the system  (\ref{autonomousb}) for $\xi=1$. Without lack of generality we use $\alpha=0.5.$  Observe that $B(\pm \xi)$ and $C(\pm \xi)$ are local sources and the point $F$ located at the curve $G$ is a local attractor. The curve K is stable, but not asymptotically stable. The scalar field-dominated solution $D(\pm\xi)$ are of saddle type, as well as the rest of the (curves of) fixed points. The scaling solutions $E(\pm\xi)$ do not exist. } }
\end{center}
\end{figure*}

\begin{figure*}
\begin{center}
\begin{tabular}{cc}
\includegraphics[scale=0.5]{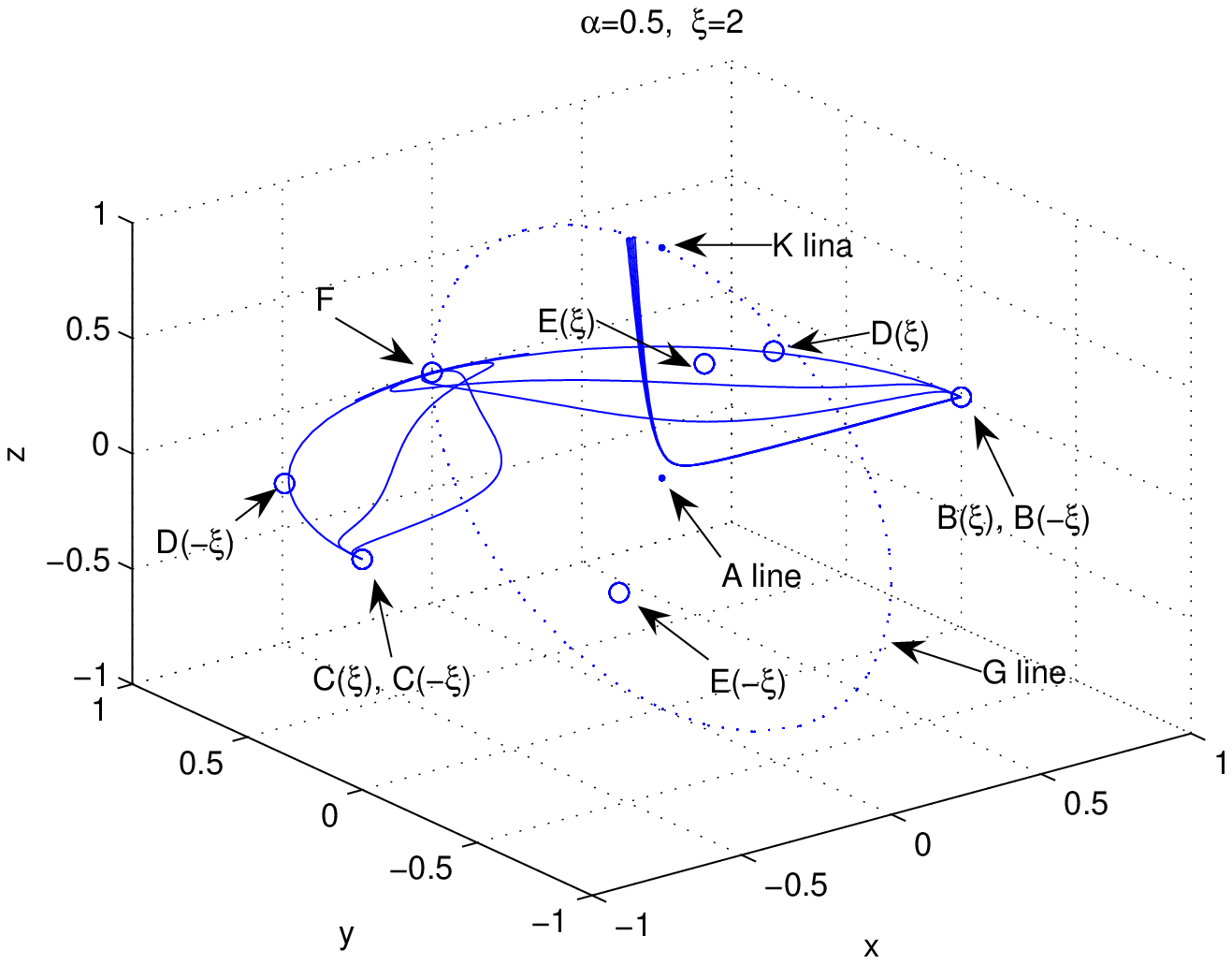} &
\includegraphics[scale=0.5]{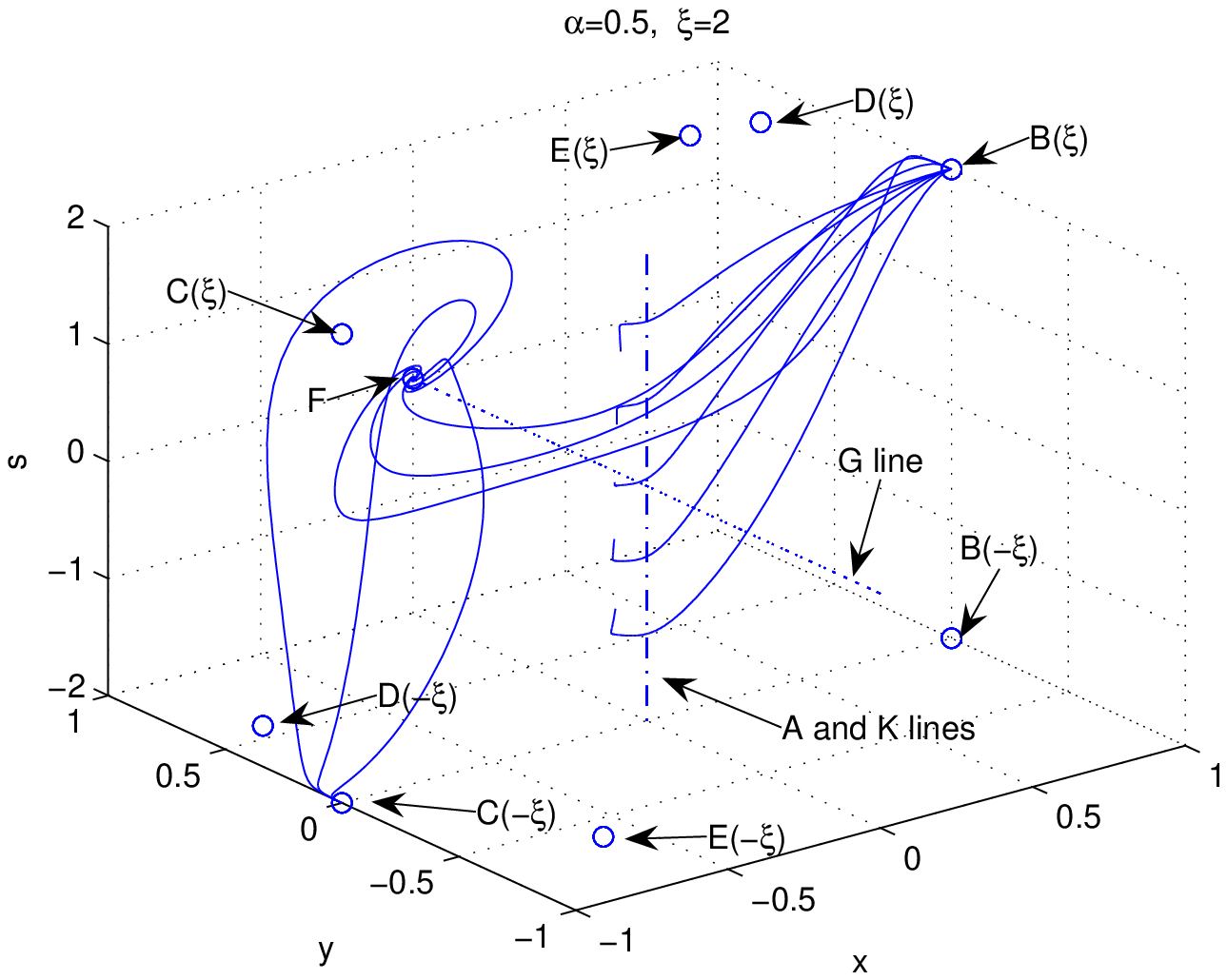} \\
\includegraphics[scale=0.5]{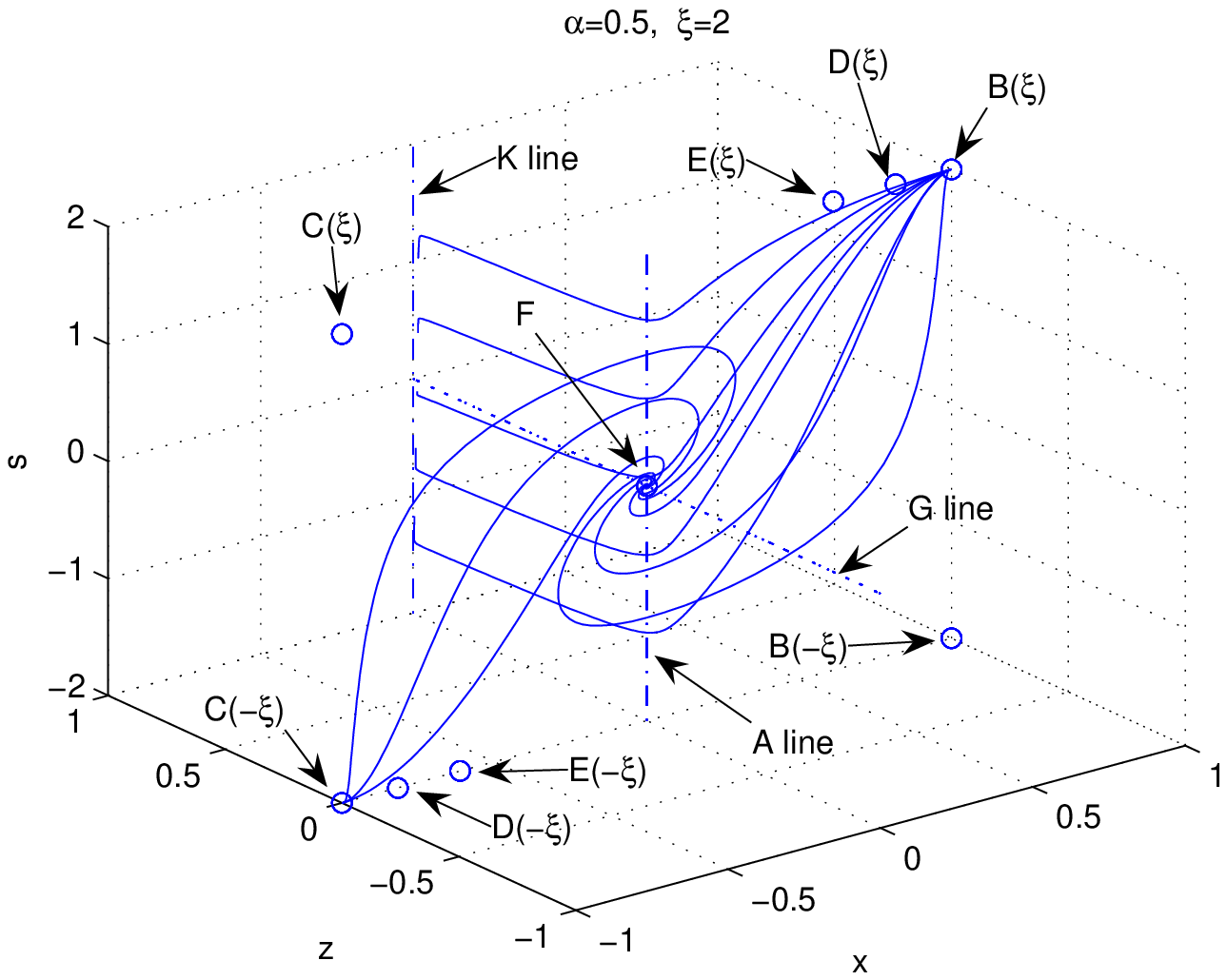} &
\includegraphics[scale=0.5]{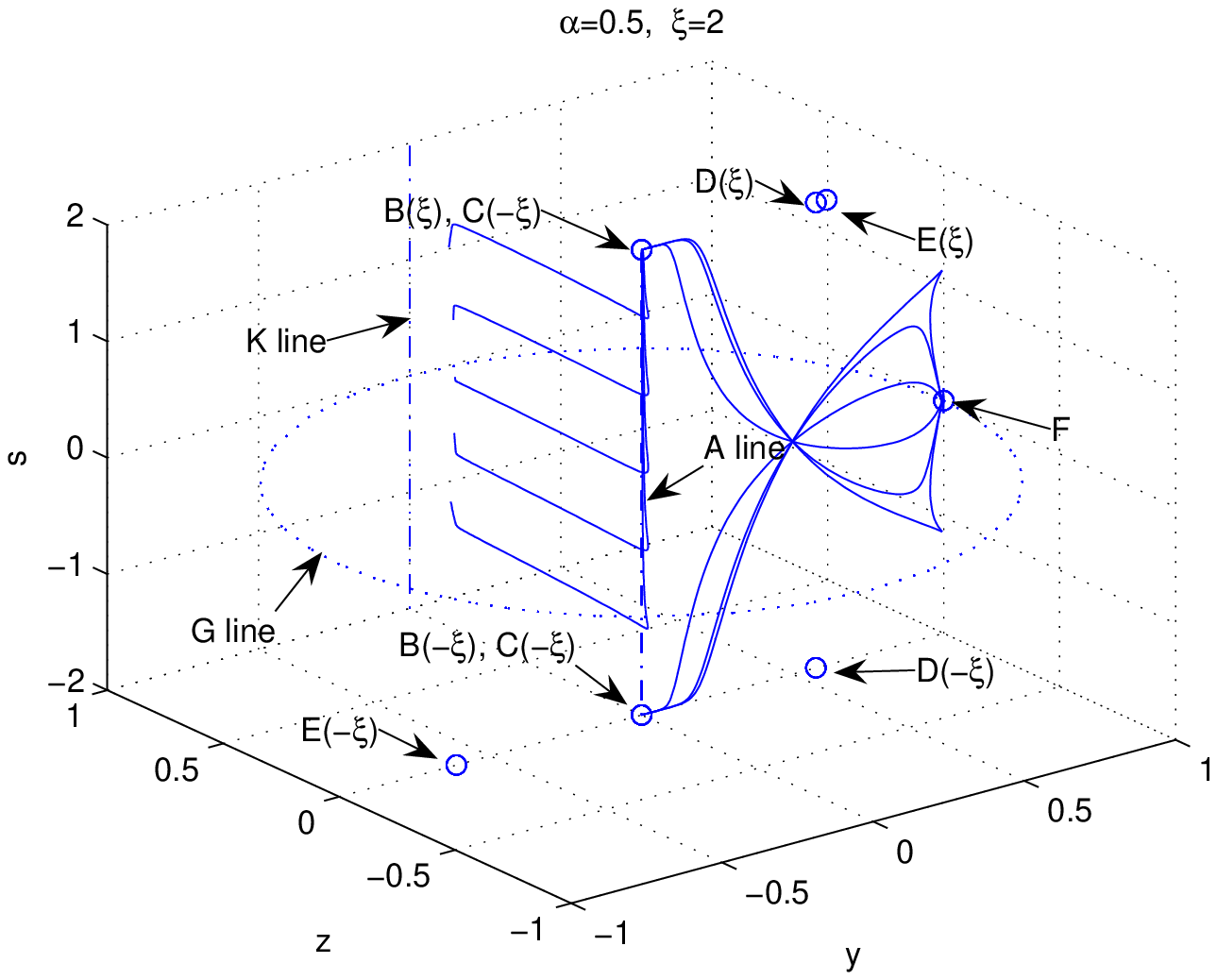}
\end{tabular}
\caption{\label{Fig5} \it{Projections of some orbits in the phase space of the system  (\ref{autonomousb}) for $\xi=2$. Without lack of generality we use $\alpha=0.5.$ Note that $B(\pm \xi)$ and $C(\pm \xi)$ are local sources and the point $F$ located at the curve $G$ is a local attractor.  The curve K is stable, but not asymptotically stable. The scalar field-dominated solution $D(\pm\xi)$ and the scaling solution $E(\pm\xi)$ are of saddle type, as well as the rest of the (curves of) fixed points. }}
\end{center}
\end{figure*}

\begin{figure*}
\begin{center}
\begin{tabular}{cc}
\includegraphics[scale=0.5]{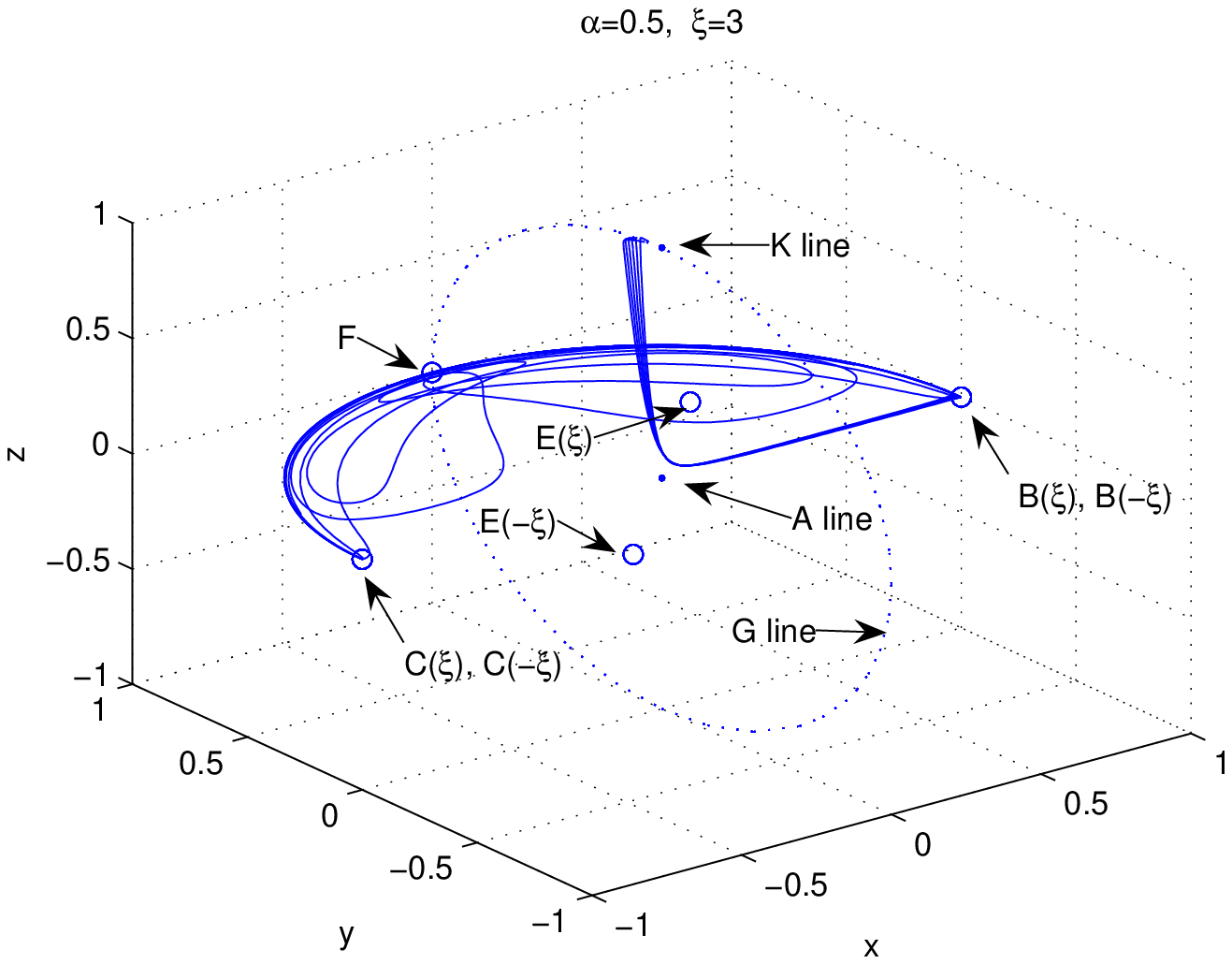} &
\includegraphics[scale=0.5]{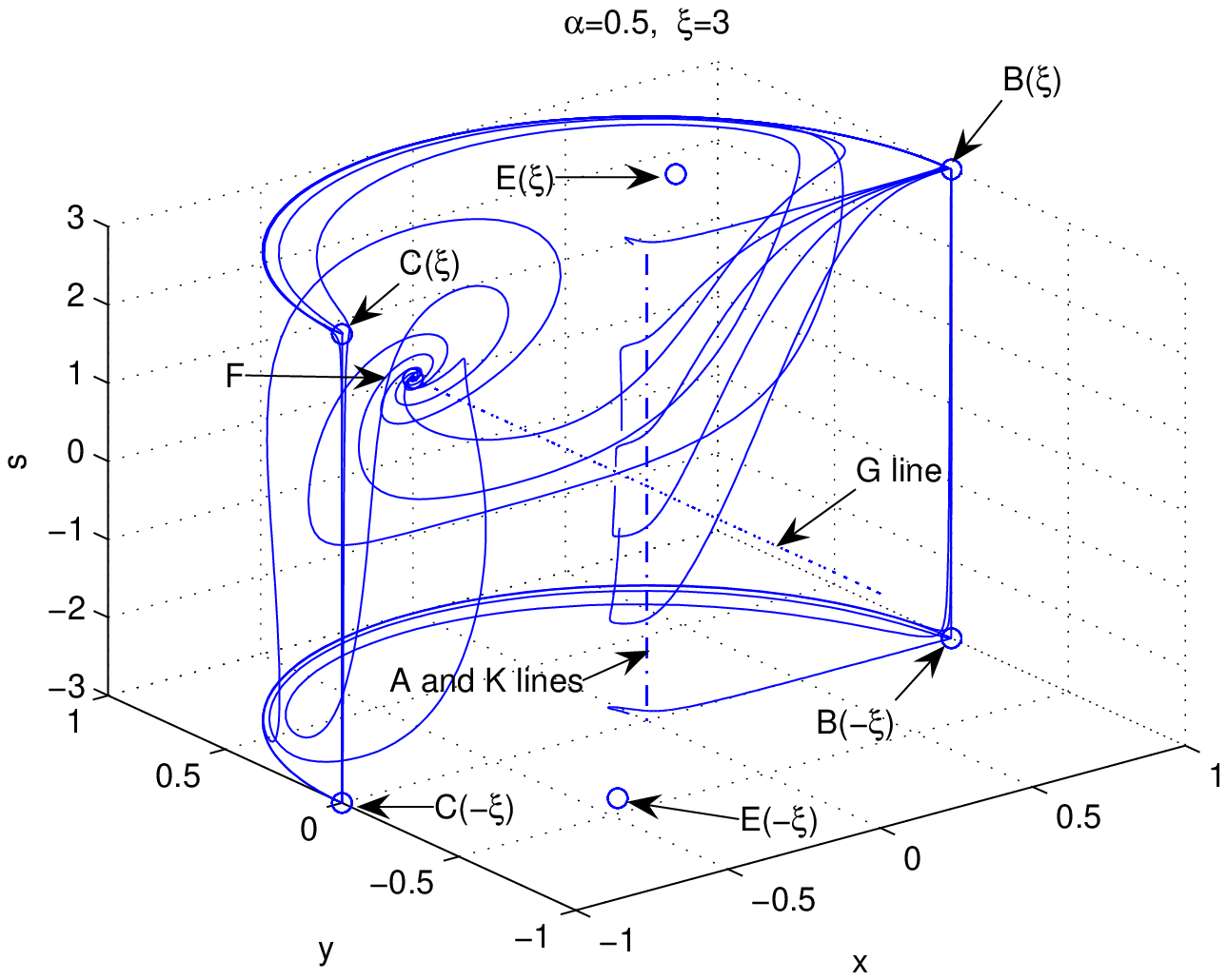} \\
\includegraphics[scale=0.5]{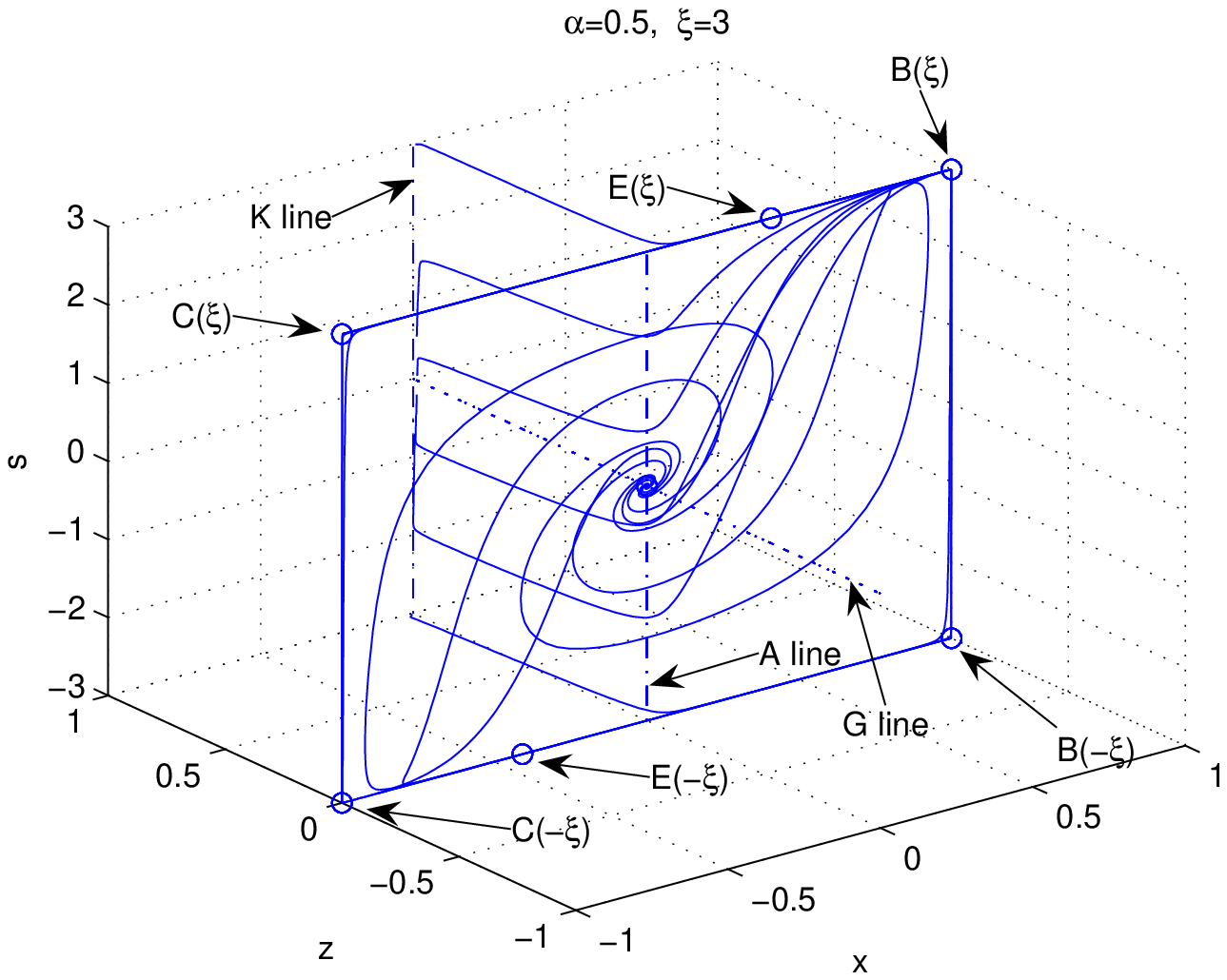} &
\includegraphics[scale=0.5]{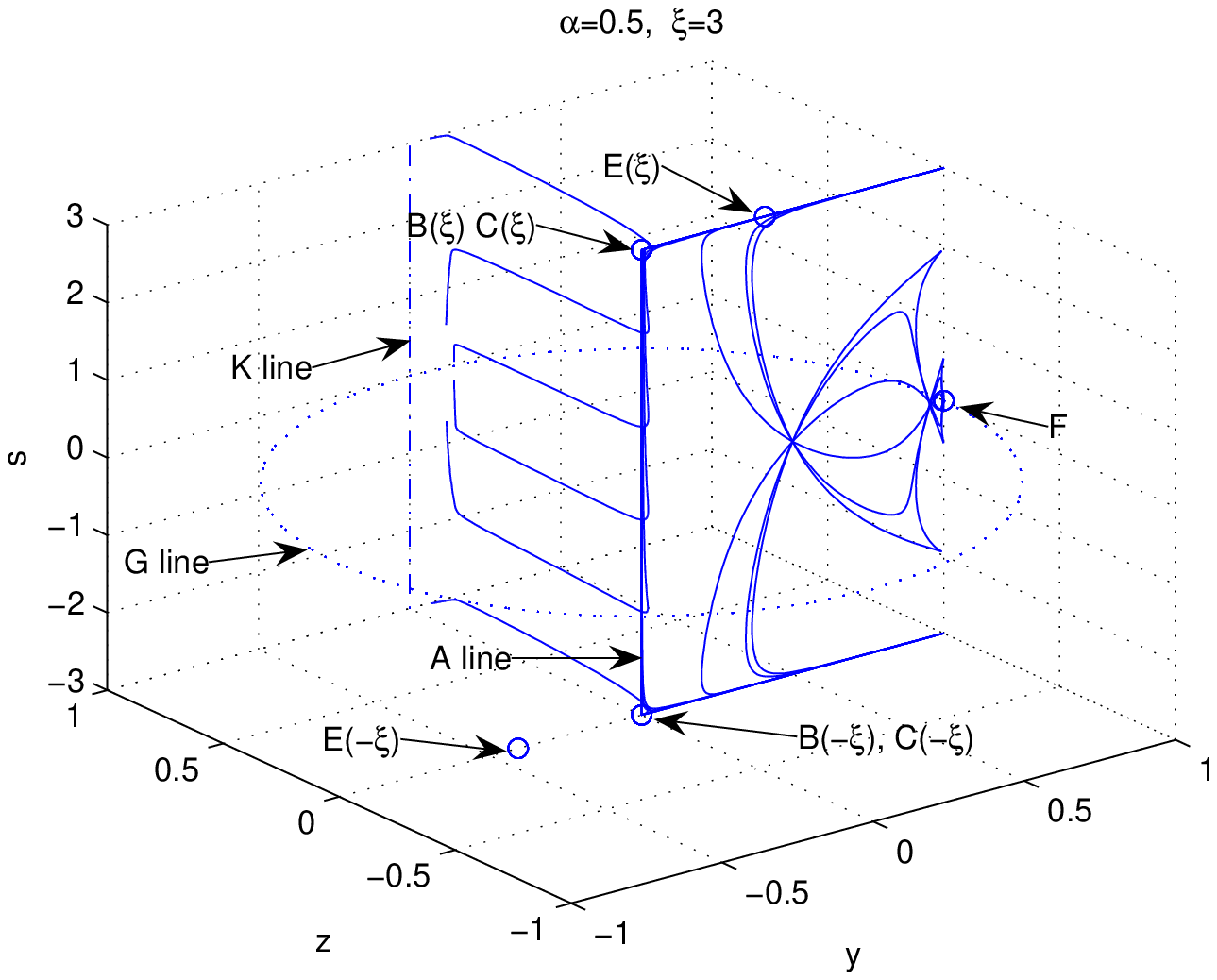}
\end{tabular}
\caption{\label{Fig6} \it{Projections of some orbits in the phase space of the system  (\ref{autonomousb}) for $\xi=3$. Without lack of generality we use $\alpha=0.5.$  $B(\pm \xi)$ and $C(\pm \xi)$ are local sources and the point $F$ located at the curve $G$ is a local attractor.  The curve K is stable, but not asymptotically stable. The scalar field-dominated solution $D(\pm\xi)$ and the scaling solution $E(\pm\xi)$ are of saddle type, as well as the rest of the (curves of) fixed points. }}
\end{center}
\end{figure*}

To finish this section let us discuss some numerical simulations:

\begin{itemize}

\item In the Fig. \ref{Fig4} are displayed the projections of some orbits in the phase space of the system  (\ref{autonomousb}) for $\alpha=0.5, \xi=1.$  $B(\pm \xi)$ and $C(\pm \xi)$ are local sources and the point $F$ located at the curve $G$ is a local attractor. The curve K is stable, but not asymptotically stable. The scalar field-dominated solution $D(\pm\xi)$ are of saddle type, as well as the rest of the (curves of) fixed points. The scaling solutions $E(\pm\xi)$ do not exist.

\item In the Fig. \ref{Fig5} are displayed some projections of some orbits in the phase space of the system  (\ref{autonomousb}) for $\alpha=0.5, \xi=2.$ Note that $B(\pm \xi)$ and $C(\pm \xi)$ are local sources and the point $F$ located at the curve $G$ is a local attractor.  The curve K is stable, but not asymptotically stable. The scalar field-dominated solution $D(\pm\xi)$ and the scaling solution $E(\pm\xi)$ are of saddle type, as well as the rest of the (curves of) fixed points.

\item In the Fig. \ref{Fig6} are presented the projections of some orbits in the phase space of the system  (\ref{autonomousb}) for $\alpha=0.5, \xi=3.$ Observe that $B(\pm \xi)$ and $C(\pm \xi)$ are local sources and the point $F$ located at the curve $G$ is a local attractor.  The curve K is stable, but not asymptotically stable. The scalar field-dominated solution $D(\pm\xi)$ and the scaling solution $E(\pm\xi)$ are of saddle type, as well as the rest of the (curves of) fixed points.

\end{itemize}

Summarizing, for arbitrary potentials, using the numerical simulations to support our conjectures, and employing analytical tools as the main proof,  we have corroborated  that $F$ is a late-time attractor which is contained in the curve $G.$ The curve of fixed points $G$ and $K$ are stable but not asymptotically stable. The numerical simulations suggest that the early time attractors are:
 \begin{itemize}
 \item $B(\xi)$ or $C(-\xi)$ for $0<\xi<\sqrt{6};$
 \item $B(-\xi)$ or $C(\xi)$ for $-\sqrt{6}<\xi<0.$
\end{itemize}
The rest of the (curves of) critical points are saddle points.

\section{Final Remarks}\label{Final_rem}

In the present work we have studied the thawing dark energy
scenarios with the Chaplygin gas as the other matter content of the universe and different kinds of self-interacting potentials for the scalar field. First, we obtain the exact solution for the cosmological equation of our model in terms of the elliptic function of the first and second kind and we obtain the right limit in the case of dust matter. Second, we used the cosmological model for the exponential potential for the scalar field and also we perform the dynamical systems analysis and we characterize the phase space of this system. We found the critical point of this system and also we studied the stability of this point. The main characteristic of this phase space are:
\begin{itemize}

\item The critical point $A$ is always a saddle. It represents cosmological solutions dominated by the Chaplygin gas mimicking dust, this solution correlates with the transient matter dominated epoch of the universe.

\item The critical points $B$ and $C$ corresponding to stiff solutions are always unstable. $B$ ($C$ resp.) is a local source for $\lambda>-\sqrt{6}$ ($\lambda<\sqrt{6}$, resp.), otherwise  they are saddles.

\item The usual quintessence points $D$ (scalar field-dominated solution) and $E$ (the usual scaling scalar field matter solution) \cite{Copeland:1997et} cannot be the late time attractors due to the presence of a GCG with $\alpha>0$ in the background. However, in the limit $A\rightarrow 0,$ when the Chaplygin gas behaves as dust and the $z$-variable is not required, we recover the standard quintessence scenario \cite{Copeland:1997et}. This is a crucial difference with respect to previous works in the literature.

\item For a constant potential (exponential with $\lambda=0$), the solution $F$  represent a de Sitter solution which is stable but not asymptotically stable (see the Appendix \ref{A.1}).
\item  For a constant potential (exponential with $\lambda=0$), the curve of critical points $G$ is stable but not asymptotically stable (see details in the Appendix \ref{appendixA}).
\item For an exponential potential ($\lambda\neq 0$),  $K$ is asymptotically stable (see details in the Appendix \ref{appendixA}).

\end{itemize}

Finally for analyzing general potential $V(\phi)$  we used the ``Method of $f$-devisers" and we obtained the critical point for this dynamical system and also we characterized the phase space and we studied the stability of the critical point. In this case, the principal characteristic of this phase space are:
\begin{itemize}

\item For arbitrary potentials, the curve of critical points $A$, representing dust solutions is always a saddle. This solutions correlated with the transient matter dominated epoch of the universe.

\item Considering $s^*$  such that $f(s^*)=0.$ For these $s$-values, the solutions $B(s^*)$ and $C(s^*)$ are past attractors or saddle points under the same conditions of the standard quintessence scenario \cite{Copeland:1997et} with the identification $s^*\equiv \lambda$ (see table \ref{crit2}).

\item For the same $s$-values, the standard quintessence solutions  $D(s^*), E(s^*)$ are saddle points (we are considering $\alpha>0$). This is the main difference here with respect the results for the standard quintessence scenario \cite{Copeland:1997et}. In the dust limit $A\rightarrow 0,$ the standard quintessence scenario is recovered \cite{Copeland:1997et} since the $z$-variable is not required anymore in the dynamics.

\item  For arbitrary potentials and provided $f(0) > 0$,  $F$ (contained in the curve $G$) is an attractor  and for  $f(0)<0,$  it is a saddle (see Appendix \ref{A.2}).

\item For arbitrary potentials and provided $f(0) > 0$,  the
curve of fixed points $G$ is stable but not asymptotically stable (see appendix \ref{appendixB}).

\item For arbitrary potentials, the curve $K$ is stable but not asymptotically stable (see appendix \ref{appendixB}).

\end{itemize}

\begin{acknowledgments}
This work was funded by Comisi\'on Nacional de Ciencias y Tecnolog\'{\i}a through FONDECYT
 Grants 1110230 (SdC and RH), 1130628 (RH and SdC), and 1110076 (SdC and JS)  and by DI-PUCV
 Grant 123710 (SdC), 123724 (RH) and 123713 (JS). CL is supported by Grant UTA MAYOR 2013-2014 and GL
 was supported by PUCV through Proyecto DI Postdoctorado 2013.
CRF was supported by Ministerio de Educaci\'on Superior (MES) of Cuba. GL wish to thanks to his colleagues at Instituto de F\'{\i}sica, Pontificia Universidad de Cat\'{o}lica de Valpara\'{\i}so for their warm hospitality during the completion of this work.
\end{acknowledgments}

\begin{appendix}

\section{Stability analysis of the pure de Sitter solution}\label{A}

In this appendix we introduce local coordinates for analyzing the stability of the pure de Sitter solution given by the fixed point $F.$

\subsection{Exponential potential}\label{A.1}

For the stability analysis of the point $F$ we introduce the local coordinates
\begin{equation}
\left\{x,1-y,z\right\} =\left\{\hat{x}, \hat{y},\hat{z}\right\} \epsilon +{\cal O}(\epsilon)^2,
\end{equation}
where $\epsilon \ll 1,$ and $\hat{y}\geq 0,\hat{z}\geq 0.$

Then the evolution of the linear perturbations is given by the equations
\begin{align}\label{eqvolFa}
&\hat{x}'= -3\hat{x},\nonumber\\
&\hat{y}'=-3 \hat{y}+\frac{3}{2} \hat{z},\nonumber\\
&\hat{z}'=3 \alpha\hat{z} -\frac{3\alpha\hat{z}^2}{2\hat{y}}.
\end{align}
The system  \eqref{eqvolFa} admits the first integral $\hat{z} \hat{y}^\alpha=c_1,$ where $c_1$ is an integration constant.
Thus we can study the reduced system
\begin{align}\label{eqvolFareduced}
&\hat{x}'= -3\hat{x},\nonumber\\
&\hat{y}'=-3 \hat{y}+\frac{3}{2} c_1 \hat{y}^{-\alpha}.
\end{align}
The system \eqref{eqvolFareduced} admits the solution passing by $\left(\hat{x}_0, \hat{y}_0\right)$ at time $\tau=0$ given by
\begin{align}
&\hat{x}(\tau)=\hat{x}_0 e^{-3\tau},\nonumber\\
&\hat{y}(\tau)=2^{-\frac{1}{\alpha +1}} \left(c_1-e^{-3 (\alpha +1) \tau } \left(c_1-2 \hat{y}_0^{\alpha +1}\right)\right)^{\frac{1}{\alpha
   +1}},
\end{align}
where $c_1=\hat{z}_0 \hat{y}_0^\alpha,$ $z_0=z(0).$
Observe that $c_1\ll 1$ as far as $y_0$ and $z_0$ are small enough.
We have $(\hat{x},\hat{y},\hat{z})\rightarrow\left(0,2^{-\frac{1}{\alpha +1}} \left(\hat{z}_0 \hat{y}_0^\alpha\right)^{\frac{1}{\alpha +1}},2^{\frac{\alpha }{\alpha +1}} \left(\hat{z}_0 \hat{y}_0^\alpha\right)^{\frac{1}{\alpha +1}}\right),$ as $\tau\rightarrow \infty.$ Thus, for a given $\delta>0,$ and $\alpha>0,$ it is possible to choose an initial state such that $ z_0 y_0^{\alpha }<2\times \left(\frac{\delta}{5}\right)^{\frac{\alpha +1}{2}},$ which give a final state in a $\delta$-neighborhood of the origin. This implies the stability, but not the asymptotic stability of $F.$

\subsection{Arbitrary potential}\label{A.2}

For the stability analysis of the point $F$ we introduce the local coordinates
\begin{equation}
\left\{x,1-y,z,s\right\} =\left\{\hat{x}, \hat{y},\hat{z},\hat{s}\right\} \epsilon +{\cal O}(\epsilon)^2,
\end{equation}
where $\epsilon \ll 1,$ and $\hat{y}\geq 0,\hat{z}\geq 0.$
Then the evolution of the linear perturbations is given by the equations
\begin{align}\label{eqvolFb}
&\hat{x}'= -3\hat{x}+\sqrt{\frac{3}{2}} \hat{s},\nonumber\\
&\hat{y}'=-3 \hat{y}+\frac{3}{2} \hat{z},\nonumber\\
&\hat{z}'=3 \alpha\hat{z} -\frac{3\alpha\hat{z}^2}{2\hat{y}}, \nonumber\\
&\hat{s}'=-\sqrt{6} \hat{x} f(0).
\end{align}
The system  \eqref{eqvolFb} admits the first integral $\hat{z} \hat{y}^\alpha=c_1,$ where $c_1$ is an integration constant.
Thus we can analysis the reduced system
\begin{align}\label{eqvolFbreduced}
&\hat{x}'= -3\hat{x}+\sqrt{\frac{3}{2}} \hat{s},\nonumber\\
&\hat{y}'=-3 \hat{y}+\frac{3}{2} c_1 \hat{y}^{-\alpha}, \nonumber\\
&\hat{s}'=-\sqrt{6} \hat{x} f(0).
\end{align}

The system \eqref{eqvolFbreduced} admits the solution passing by $\left(\hat{x}_0, \hat{y}_0,\hat{s}_0\right)$ at time $\tau=0$ given by

\begin{align} &\hat{x}(\tau)=\frac{\sqrt{\frac{3}{2}} \hat{s}_0 e^{-3 \tau /2} \sinh (\xi  \tau )}{\xi }+\nonumber \\
    &+ \frac{e^{-3 \tau /2} \hat{x}_0 (2 \xi  \cosh (\xi  \tau )-3
   \sinh (\xi  \tau ))}{2 \xi }, \nonumber \\
    &\hat{y}(\tau)=2^{-\frac{1}{\alpha +1}} \left(c_1-e^{-3 (\alpha +1) \tau } \left(c_1-2 \hat{y}_0^{\alpha +1}\right)\right)^{\frac{1}{\alpha
   +1}},\nonumber\\
    & \hat{s}(\tau)=\frac{\hat{s}_0 e^{-3 \tau /2} (3 \sinh (\xi  \tau )+2 \xi  \cosh (\xi  \tau ))}{2 \xi }+\nonumber \\
    &+\frac{\left(4 \xi ^2-9\right) e^{-3 \tau /2}
   \hat{x}_0 \sinh (\xi  \tau )}{2 \sqrt{6} \xi },
\end{align}
where $\beta=\frac{1}{2} \sqrt{9-12 f(0)}.$

For the choice $\beta^2<\frac{9}{4},$ i.e., $f(0)>0,$ $(\hat{x},\hat{y},\hat{z},\hat{s})\rightarrow\left(0,2^{-\frac{1}{\alpha +1}} \left(\hat{z}_0 \hat{y}_0^\alpha\right)^{\frac{1}{\alpha +1}},2^{\frac{\alpha }{\alpha +1}} \left(\hat{z}_0 \hat{y}_0^\alpha\right)^{\frac{1}{\alpha +1}},0\right),$ as $\tau\rightarrow \infty.$
For $\beta=\pm \frac{3}{2},$ i.e., for $f(0)=0,$  $(\hat{x},\hat{y},\hat{z},\hat{s})\rightarrow\left(\frac{\hat{s}_0}{\sqrt{6}},2^{-\frac{1}{\alpha +1}} \left(\hat{z}_0 \hat{y}_0^\alpha\right)^{\frac{1}{\alpha +1}},2^{\frac{\alpha }{\alpha +1}} \left(\hat{z}_0 \hat{y}_0^\alpha\right)^{\frac{1}{\alpha +1}},\hat{s}_0\right),$ as $\tau\rightarrow \infty.$   Combining the above arguments we obtain that for $f(0)\geq 0,$ $F$ is stable, but not asymptotically stable.
For $\beta^2>\frac{9}{4},$ i.e., $f(0)<0,$ the perturbation values $\hat{x}$ and $\hat{s}$ diverges, and $(\hat{y},\hat{z})\rightarrow\left(2^{-\frac{1}{\alpha +1}} \left(\hat{z}_0 \hat{y}_0^\alpha\right)^{\frac{1}{\alpha +1}},2^{\frac{\alpha }{\alpha +1}} \left(\hat{z}_0 \hat{y}_0^\alpha\right)^{\frac{1}{\alpha +1}}\right),$  as $\tau\rightarrow
+\infty.$ Thus, $F$ is a saddle for $f(0)<0$.

\section{Center Manifold calculations for an scalar field with exponential potential}\label{appendixA}

For analyzing the stability of the curve of critical points $G$ (which exists only for $\lambda=0$) we introduce the new coordinates
\begin{align}
&u_1=\frac{y_c \left(2 y \left(y_c^2-\alpha -1\right)+y_c \left(-y_c^2+z+2 \alpha +1\right)\right)}{\alpha
   +1},\nonumber\\
& v_1=x,\nonumber\\
& v_2=-\frac{\left(y_c^2-\alpha -1\right) \left(2 y y_c-y_c^2+z-1\right)}{\alpha +1},
\end{align} which are referred to an arbitrary point at $G$ with coordinates $(0, y_c,1-y_c^2).$
Applying the procedure, we find that the center manifold is given by the graph
\begin{align}
&\left\{(u_1,v_1,v_2): v_1={\cal O}(5),  v_2=\frac{u_1^2 \left(y_c^2-\alpha -1\right)}{4 y_c^2 (\alpha +1)}+\right. \nonumber  \\ & \left.  +\frac{u_1^3 \left(y_c^2-\alpha -1\right)}{8 y_c^2 (\alpha
   +1)^2} +\frac{5 u_1^4 \left(y_c^2-\alpha -1\right)}{64 y_c^2 (\alpha +1)^3}+\right. \nonumber  \\ & \ \ \ \ \ \ \ \ \ \ \left. +{\cal O}(5), |u_1|<\delta\right\},
\end{align}
where $\delta$ is a small enough constant, and ${\cal O}(5)$ denotes terms of five order in the vector norm.
The dynamics on the center manifold is governed by the equation
$$u_1'={\cal O}(5).$$ From this follows that $G$ is stable but not asymptotically stable.
The center manifold of $K$ is given by the approximated graph
\begin{align}
& \left\{(x,y,z): x=\frac{u^2 \lambda }{\sqrt{6}}+{\cal O}(5), y=u, \right. \nonumber  \\ & \left. z=1-\frac{u^4 (\alpha -1) \lambda ^2}{6 (\alpha +1)}-u^2+{\cal O}(5), |u|<\delta\right \},
\end{align}
where ${\cal O}(5)$ denote terms of order $5$ with respect the vector norm.

The dynamics on the center manifold of $K$ is dictated by the gradient-like equation
\begin{equation}
u'=-\frac{u^3 \lambda ^2}{2}+{\cal O}(5).
\end{equation}

Since the origin is a degenerated minimum of the potential $U(u)=\frac{u^4 \lambda ^2}{8}$ follows the stability of $K.$

\section{Center manifold calculations for an scalar field with arbitrary potential}\label{appendixB}

For study the stability of $G$ we resort the the Center Manifold Theory.
Let us assume that $0<f(0)\leq\frac{4}{3}.$ Then, introducing the new variables
\begin{align}
&u_1=\frac{y_c \left(2 y \left(y_c^2-\alpha -1\right)+y_c \left(-y_c^2+z+2 \alpha +1\right)\right)}{\alpha +1},\nonumber\\
&v_1=-\frac{\left(y_c^2-\alpha -1\right) \left(2 y
   y_c-y_c^2+z-1\right)}{\alpha +1},\nonumber\\
&v_2=   \frac{s \left(\sqrt{6-8 f(0) y_c^2}-\sqrt{6}\right)+4 f(0) x}{2 \sqrt{6-8 f(0) y_c^2}},\nonumber\\
&v_3=\frac{s \sqrt{6-8 f(0) y_c^2}-4 f(0) x+\sqrt{6} s}{2
   \sqrt{6-8 f(0) y_c^2}},
\end{align}
and applying the procedure, we find that the center manifold is given by the graph
\begin{align}
&\left\{(u_1,v_1,v_2,v_3): v_1=g(u_1)+{\cal O}(5),  \right. \nonumber  \\ & \left. v_2={\cal O}(5), v_3={\cal O}(5), |u_1|<\delta\right\},
\end{align}
where $g(u_1)=\frac{u_1^2 \left(5 u_1^2+8 u_1 (\alpha +1)+16 (\alpha +1)^2\right) \left(y_c^2-\alpha -1\right)}{64 y_c^2 (\alpha +1)^3},$ $\delta$ is a small enough constant, and ${\cal O}(5)$ denotes terms of five order in the vector norm.

The dynamics on the center manifold is governed by the equation
$$u_1'={\cal O}(5).$$ Form this follows that $G$ is stable but not asymptotically stable.

For analyzing the case of complex eigenvalues ($f(0)>\frac{4}{3}$), we can introduce the new variables
\[V_2=\frac{v_2+v_3}{2},\; V_3=\frac{v_2-v_3}{2i},\] for deriving the real Jordan form of the Jacobian. The procedure is straightforward and the result is the same.

For analyzing the stability of the curve of critical points $K$ we proceed as follows.

Let us assume $s_c\neq 0.$ Introducing the new variables
\begin{equation}\label{newvars}
u_1=s-s_c-\sqrt{\frac{3}{2}}x f(s_c), \, u_2=y,\, v_1=\sqrt{\frac{3}{2}}x f(s_c),\, v_2=z,
\end{equation}
and applying the procedure, we find that the center manifold is given by the graph
\begin{align}
& \left\{(u_1,u_2,v_1,v_2): v_1=\frac{1}{3}s_c u_2^2+\frac{1}{3}u_1 u_2^2 f(s_c)+{\cal O}(4), \right. \nonumber\\   & \left.  v_2=-u_2^2+{\cal O}(4), u_1^2+u_2^2<\delta\right\},
\end{align}
where $\delta$ is a small enough constant, and ${\cal O}(4)$ denotes terms of fourth order in the vector norm.
But since $v_2\equiv z\geq 0,$ it follows that $u_2$ should be zero. Thus, the center manifold of the origin is
\begin{align}
& \left\{(u_1,u_2,v_1,v_2):u_2=0, v_1={\cal O}(4), \right. \nonumber\\   & \left.  v_2={\cal O}(4), u_1^2+u_2^2<\delta\right\}.
\end{align}

The dynamics on the center manifold is governed by the equations
\begin{align}\label{centerH}
&u_1'={\cal O}(4).
\end{align}

From this fact follows the stability (but not the asymptotic stability) of the center manifold of the origin, thus, follow the the stability (but not asymptotic stability) of $K$.
\end{appendix}

\end{document}